\documentclass[authoryear,12pt]{elsarticle}
\makeatletter
   \def\ps@pprintTitle{%
      \let\@oddhead\@empty
      \let\@evenhead\@empty
	  \def\@oddfoot{\centerline{\thepage}}%
      \let\@evenfoot\@oddfoot
   }
\makeatother

\usepackage{amssymb,amsmath,stmaryrd,footmisc}
\usepackage{esint}
\usepackage{placeins}
\usepackage{xcolor}
\usepackage{epstopdf}
\usepackage{natbib}
\usepackage{nohyperref}
\usepackage{graphicx}
 \usepackage{setspace}
\usepackage{color}
\usepackage{verbatim}
 \usepackage[usenames,dvipsnames]{pstricks}
 \usepackage{epsfig}
 \usepackage{pst-grad} 
 \usepackage{pst-plot} 

\usepackage{etoolbox}
\usepackage{mathtools}
\preto\subequations{\ifhmode\unskip\fi}

\usepackage{physics}
\usepackage{amsmath}

\DeclareMathOperator*{\argmin}{arg\,min}

\newlength{\tleft}
 \newlength{\tright}
\def\nten#1{\mathbf{#1}}

\renewcommand{\vec}[1]{\mathbf{#1}}

\def\dD0{\mathcal{\partial D}_0}
\def\D0{\mathcal{D}_0}

\def\dV0{\dd{V_0}}

\def\dA0{\dd{A_0}}

\def\jump#1{\llbracket #1 \rrbracket}

\journal{Journal of the Mechanics and Physics of Solids}
\usepackage{cleveref}

\begin{document}

\begin{frontmatter}

\title{Shear shock evolution in incompressible soft solids}

\author[label1]{S. Chockalingam}

\author[label2,label3]{T. Cohen\corref{cor1}}
\cortext[cor1]{Corresponding author: talco@mit.edu}

\address[label1]{Massachusetts Institute of Technology, Department of Aeronautics and Astronautics, Cambridge, MA, 02139, USA}
\address[label2]{Massachusetts Institute of Technology, Department of Civil and Environmental Engineering, Cambridge, MA, 02139, USA}
\address[label3]{Massachusetts Institute of Technology, Department of Mechanical Engineering, Cambridge, MA, 02139, USA}



\begin{abstract}
Nonlinear evolution of shear waves into shocks in incompressible elastic materials is investigated using the framework of large deformation elastodynamics, for a family of loadings and commonly used hyperelastic material models. Closed form expressions for the shock formation distance are derived and used to construct non-dimensional phase maps that determine regimes in which a shock can be realized. These maps reveal the sensitivity of shock evolution to the amplitude, shape, and ramp time of the loading, and to the elastic material parameters. In light of a recent study (Espindola et al., 2017), which hypothesizes that shear shock formation could play a significant role in Traumatic Brain Injury (TBI), application to brain tissue is considered and it is shown that the size matters in TBI research. Namely, for realistic loadings, smaller brains are less susceptible to formation of shear shocks. Furthermore, given the observed sensitivity to the imparted waveform and the constitutive properties, it is suggested that the non-dimensional maps can guide the design of protective structures by determining the combination of loading parameters, material dimensions, and elastic properties that can avoid shock formation.
\end{abstract}


\end{frontmatter}


\section{Introduction}

\noindent Solids are capable of transmitting mechanical shear waves, which can evolve into shocks depending on the nonlinearity of the shear response. Generating such shear shocks in stiff materials, such as metals, would require extreme strain rates at high stress and would lead to catastrophic damage and failure by other competing processes before shock wave phenomena can be observed  \citep{marchand1988experimental,mercier1998steady}. By contrast, even weak impacts can generate shear shocks within small distances in soft solids, owing to their low shear moduli and significantly higher nonlinearity in shear  \citep{catheline2003observation}. Many soft solids including biological tissues typically have a shear modulus that is orders of magnitude smaller than the bulk modulus and can hence be assumed to be incompressible. This paper concerns the nonlinear evolution of shear waves into shear shocks in such materials.

Nonlinear shear waves have been studied extensively by both the continuum mechanics and nonlinear acoustics communities. Propagation of finite amplitude plane shear waves and shear shocks were investigated by \cite{chu1964finite,chu1967transverse}, in incompressible isotropic elastic materials. Therein, the condition for shear shock formation was established and the shear loading problem of an incompressible elastic half space was studied using the method of characteristics. \cite{collins1966one,collins1967propagation} extended the investigation of wave propagation to materials with transverse isotropy and allowed for transverse displacements in two directions. 
  \cite{davison1966propagation} studied nonlinear shear waves and propagation of shock waves that are generated by impact loadings in compressible hyperelastic materials wherein accounting for coupled longitudinal motion is necessitated by the compressibility.  There, an instantaneous constant loading is considered whereby a shock would immediately form at the loading surface, as opposed to evolution of a smooth loading waveform into shocks as considered by \cite{chu1964finite}. Recently, \cite{ZIV201967} applied the formulation developed by \cite{davison1966propagation} to specific compressible hyperelastic material models. \cite{aboudi1973one,aboudi1974finite} developed a finite-difference based scheme to study such impact induced nonlinear waves. Solutions to the wave equations for compressible hyperelastic materials was provided by \cite{destrade2005}, including  the  possibility  of  dissipation, extending the pioneering works of \cite{Carroll1967,Carroll1974, carroll1977plane,Carroll1977_2, carroll1979reflection}.

The nonlinear evolution of shear waves has also been extensively studied in acoustics, where
it is routine to perform expansions of the strain energy density function, which is essentially an
assumption of weak nonlinearity. The paraxial approximation of small but finite\footnote{The shear response would be linear for infinitesimal displacements.}  wave amplitudes is also employed to obtain reduced wave equations. A nonlinear parabolic wave equation for shear wave beams in isotropic solids was first derived by \cite{zabolotskaya1986sound} accounting for nonlinearity, viscous dissipation and diffraction. In the absence of diffraction, the equation reduces to the Modified Burgers equation (MBE) which is similar to the Burgers equation except the quadratic nonlinearity term is replaced by a cubic one. The MBE was studied in detail by \cite{lee1987nonlinear}. An alternative expansion of the strain energy density, suitable for application to soft solids, was provided by \cite{hamilton2004separation} and used to derive equations that describe nonlinear propagation of plane shear waves for different polarizations \citep{zabolotskaya2004modeling}. The plane wave model was extended to account for diffraction by \cite{wochner2008cubic}. Recently, \cite{destrade2019} extended the formulation to any isotropic incompressible solid without having to rely on expansions of the strain energy density.

While the theoretical investigation of shear shocks in solids dates back to the 60's, the first observation of shear shocks in an elastic medium was only reported recently \citep{catheline2003observation}.  This was achieved by application of a transient elastography technique to measure the displacements induced by transverse vibrations applied on one end of a gelatin phantom sample. A more recent study by \cite{espindola2017shear} employed a similar method and demonstrated the spontaneous evolution of smooth shear waves into shear shocks in a porcine brain and reported acceleration magnifications of up to a factor of $8.5$. Hence, it was suggested that shear shock waves could be an unappreciated damage mechanism that could play a significant role in traumatic brain injury. To reliably study shear shock generation in the brain and other soft solids, it is vital to consider realistic loadings and the possibility of large deformations. However, both the experimental studies discussed, apply harmonic loading, and make use of reduced wave equations and strain energy density expansions that apply for small amplitude deformations and weak nonlinearity. 

In light of these recent studies, in this work we utilize a large deformation elastodynamics framework to study shear shock evolution using realistic constitutive relations that can capture the strongly nonlinear response of soft and biological materials, with the goal of answering the fundamental question: \textit{Can shear shocks be induced in soft materials, such as brain tissue, within the length of the impacted object, when subjected to realistic loadings at large deformations?} 
To answer the question, we conduct a parametric analysis using three different constitutive relations that capture a wide range of incompressible material response, and a family of loading waveforms. 

The paper is organized as follows: In the following section, we begin by defining our problem setting and deriving the governing equations. Then, in Sec. \ref{sec:bvp}, we solve the boundary value  problem by employing the method of characteristics, as in \cite{chu1964finite}, and derive general expressions needed to evaluate the shock formation distance. Subsequently, in Sec. \ref{sec:applied_constit}, we apply these expressions to analyze various stress responses, and arrive at closed form expressions for the shock formation distance. These expressions are used to construct non-dimensional phase maps that determine regimes in which a shock can exist and illuminate the effect of material properties and loading scenarios on shock evolution. A parallel supplementary investigation of the distance taken for realization of a given finite acceleration magnification is also carried out. Finally, in Sec. \ref{sec:Application}, we demonstrate application of our results to the problem of shear impact of the brain to answer the question  posed earlier. We provide some concluding remarks in Sec. \ref{sec:Conclusions}. 

\section{Nonlinear Shear Wave Equation and Method of Characteristics}
\smallskip

\begin{figure}
    \begin{center}
\includegraphics[width =\textwidth]{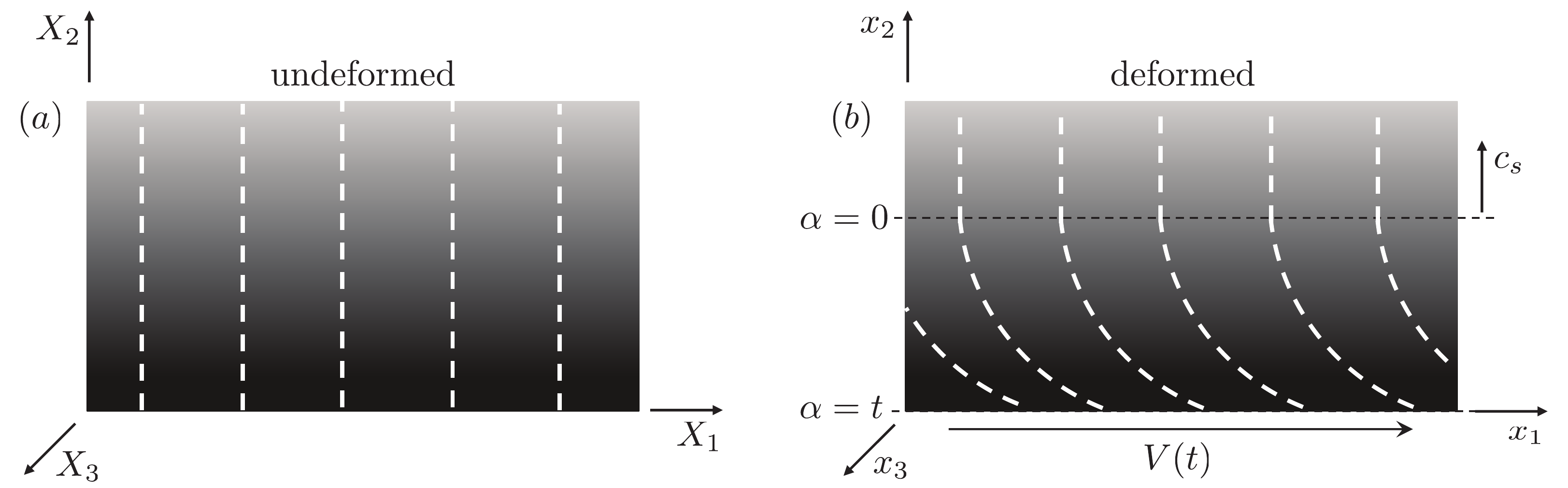}
\caption{Simple shear deformation of a homogeneous isotropic perfectly incompressible material occupying the half-space $X_2 \ge 0$.  (a) The initial undeformed configuration. (b) Deformed configuration where a continuous time varying shearing velocity $V(t)$ is applied on the surface $X_2=0$. The white dashed lines are shown to observe the shear deformation. The wavefront, $\alpha=0$ (see Sec. \ref{sec:bvp}), travels at the linear elastic shear wavespeed $c_s$ (before shock formation). } 
\label{fig:Halfspace}
\end{center}
\end{figure}

\noindent Consider a homogeneous and isotropic semi-infinite medium whose undeformed stress-free configuration is described by the Lagrangian coordinates $\vec{X}=(X_1,X_2,X_3)$, with $-\infty < X_1,X_3 < \infty$, and $X_2\ge0$, as shown in Fig. \ref{fig:Halfspace}(a). The body is subjected to shearing motion  
 by imposition of a continuous time dependent shearing velocity $V(t)$, on its surface $X_2=0$, along $X_1$ as shown in Fig. \ref{fig:Halfspace}(b). Using symmetry considerations and incompressibility, the mapping between the Lagrangian coordinates of a material point, $\vec{X}$, and its current coordinates, $\vec{x}$, at time $t$, is necessarily given by
\begin{equation}
x_1 = X_1 + u(X_2,t),\qquad x_2 = X_2,\qquad x_3 = X_3
\end{equation}
where $u(0,t) = \int_{0}^{t}V(t)\textrm{d}t$ is the shear displacement of the surface $X_2=0$. The shear strain $\gamma(X_2,t)$ is given by
\begin{equation}
\gamma(X_2,t) = \pdv{x_1}{X_2} = \pdv{u(X_2,t)}{X_2} \label{strain_def}
\end{equation}
and the particle velocity is given by 
\begin{equation}
v(X_2,t) = \pdv{u(X_2,t)}{t} \quad \textrm{where} \quad v(0,t) = V(t) \label{vel_def}
\end{equation}
\noindent The deformation gradient $\nten F$ and the  left Cauchy Green deformation tensor $\nten B$ can then be written as
\begin{equation}
\nten F = \pdv{\vec{x}}{\vec{X}} = \begin{bmatrix} 1 & \gamma & 0 \\ 0 & 1 & 0 \\ 0 & 0 & 1 \end{bmatrix} \quad , \quad \nten{B} = \nten{F}\cdot\nten{F}^T = \begin{bmatrix} \gamma^2+1 & \gamma & 0 \\ \gamma & 1 & 0 \\ 0 & 0 & 1 \end{bmatrix} 
\label{defgradientshear}
\end{equation}
and the invariants of $\nten B$ are given by $I_1 = I_2 = 3+\gamma^2$ and $J = \textrm{det}(\nten F) = 1$. 

In our analysis we consider a purely mechanical theory and restrict our attention to incompressible hyperelastic materials. The strain energy density per unit volume, $W$, can then be written in its most general form as
\begin{equation}
W = W(I_1,I_2)
\end{equation} 
Defining $W_i = \pdv{W}{I_i}$ for $i=1,2$, the first Piola-Kirchhoff stress tensor $\nten P$ and the Cauchy stress $\nten T$ are given by
\begin{equation}
\nten P =\nten{T}\nten{F}^{-T}  \quad  , \quad {\nten T}=-p~{\nten  {{\mathit{ {\nten 1}}}}}+2 \big[(W_1+I_1W_2)\nten{B}- W_2 \nten{B}^2\big] \label{stress_tensor_general}
\end{equation}  
 where $p=p(X_2,t)$ is a pressure field that arises due to the incompressibility constraint and is determined by the boundary value  problem. Substituting the deformation field \eqref{defgradientshear} in \eqref{stress_tensor_general}, we arrive at a stress state of the form\footnote{Note that different authors (ex: \cite{chu1964finite}, \cite{horgan2011simple}) might report seemingly different longitudinal stress expressions for the simple shear problem, but they all differ only by a hydrostatic term which can be absorbed into the undetermined pressure field $p$.        }
\begin{equation}
\label{Piola_incom}
 \nten P(X_2,t) = \begin{bmatrix} 
2W_1+4W_2-p & 2\gamma(W_1 + W_2) & 0\\
\gamma(-2W_2+p) & 2W_1+4W_2-p & 0\\
0 & 0 & 2W_1+4W_2+2\gamma^2W_2-p  
\end{bmatrix} \\
\end{equation}
where the stress component $P_{12}(=T_{12})$ is denoted by $\tau(\gamma) (= 2\gamma(W_1 + W_2))$ and is referred to as the shear stress.
Note that $\tau(\gamma)$ is an odd function. 

\subsection{Equation of Motion} 
\noindent While dilatational stresses are transmitted instantaneously in an incompressible elastic medium, shear stresses are transmitted at a finite velocity. As the surface $X_2=0$ is put to motion to generate deformation, a shear wave propagates into the undeformed material and its propagation is governed by balance  of linear momentum, which, in absence of body forces, reads 
\begin{equation}
\label{momentum_conserv_integ_Lagrangian}
\int_{\D0}\rho_0 \dot{\vec{v}}\rm dV_0 = \int_{\dD0}\nten{P} \vec{n}_0\rm{dA_0}
\end{equation}Here $\dot{\vec{v}}$ is the material acceleration, $\rho_0$ is the constant density, and integration is performed on a subregion $\D0$ in the reference configuration. In the absence of shocks, assuming the smoothness of fields, this balance law can be localized as 
\begin{equation}
\rm Div\ \nten P = \rho_{0} \dot{\vec v}
\label{momentum_conserv_general}
\end{equation}

\noindent Substituting \eqref{Piola_incom} into \eqref{momentum_conserv_general} gives us the following equations,
\begin{subequations}
\begin{align}
\pdv{\tau}{X_2} &= \rho_0 \pdv[2]{u}{t} \label{motion1}\\ 
\pdv{P_{22}}{X_2} &= 0 \Rightarrow \pdv{p}{X_2} = \pdv{(2W_1+4W_2)}{X_2} \label{motion2}
\end{align}
\end{subequations} 
While perfect incompressibility allows us to impose an arbitrary normal traction $\sigma_n=P_{22}$ that will be instantaneously equilibrated throughout the body (as seen from \eqref{motion2}), in a compressible material, coupling between the longitudinal and shearing deformations might become significant (see for example \cite{ZIV201967}). 

Plugging \eqref{strain_def} into \eqref{motion1} gives us the nonlinear wave equation
\begin{equation}
\pdv[2]{u}{t} = c^2\pdv[2]{u}{X_2} \quad \textrm{where} \quad c(\gamma) = \sqrt{\frac{1}{\rho_0}\pdv{\tau}{\gamma}} \label{waveeqn}
\end{equation}
where in writing the expression for the wavespeed $c(\gamma)$ the tacit assumption has been made that $\tau'(\gamma)>0$, to ensure hyperbolicity of the wave equation such that shear stresses are transmitted through waves. Note that $c(\gamma)$ is an even function. At the linear elastic limit shear waves are transmitted at the constant wavespeed $c_s$ given by
\begin{equation}
c_s = \lim\limits_{\gamma\to0} c(\gamma)=\sqrt{\frac{{\mu}}{{\rho_0}}} \label{c_s_first}
\end{equation} where $\mu$ is the linear elastic shear modulus. In terms of strain and velocity, written in \eqref{strain_def} and \eqref{vel_def},  equation \eqref{waveeqn} can be written as a system of equations,
\begin{subequations}
\begin{align}
 \pdv{v}{t} &= c^2 \pdv{\gamma}{X_2} \label{shearwave}\\
 \pdv{\gamma}{t} &= \pdv{v}{X_2}   \label{compatibility}
\end{align}\label{wavesystem}
\end{subequations}
where \eqref{compatibility} is the compatibility condition.

It should be noted that, in practice, the simple shear deformation \eqref{defgradientshear} is a non-trivial one to generate, especially in the large deformation settings,  
as it requires suitable tractions to be applied on the inclined surfaces, see \cite{destrade2012simple}, \cite{horgan2011simple}, and \cite{horgan2012boundary}. Generating a dynamic simple shear deformation in finite dimensional blocks 
would be even more tedious, requiring time and spatially varying tractions on the inclined surfaces and would not correspond to any real life shearing scenarios. In the present study we consider shearing of an elastic half-space which can be considered an approximation of a physical shearing setting with large in-plane dimensions $X_1-X_3$, such that effects from the free surfaces can be neglected in the mid-section where the deformation can be assumed to be one of simple shear.

\subsection{Shear shock condition}
\label{subsec:shock_cond} 

\begin{figure}
\begin{center}
\includegraphics[width=\textwidth]{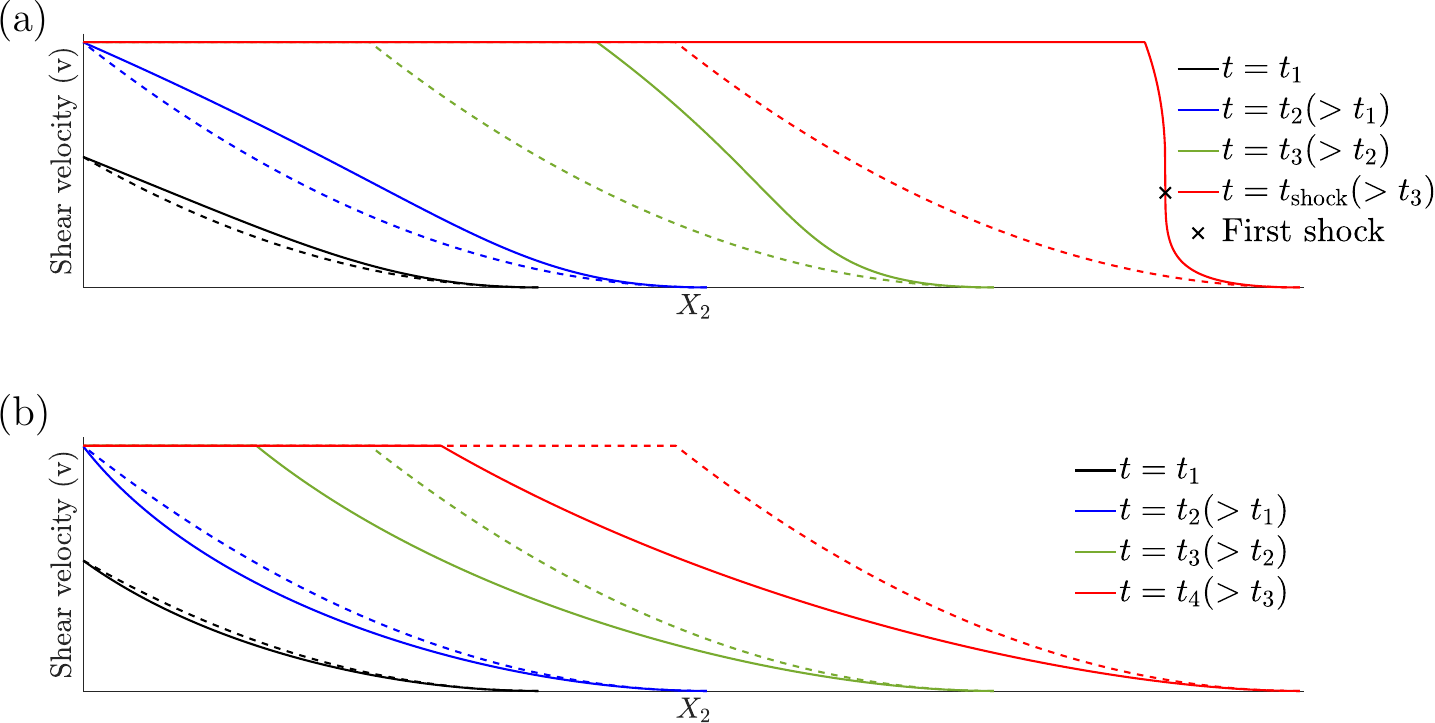}
\caption{A quadratic loading waveform ($n=2$ in \eqref{loadprofile}) nonlinearly evolving while being spatially relayed by a material with (a) Stiffening shear stress response, i.e $\tau''(|\gamma|)>0$. The figure was made for the exponential stress model \eqref{exp_stress}, and the parameters chosen were such that the loading was in the regime $m>m_\textrm{th}$ (and $n>n_c=1$) so that the first shock forms somewhere in the middle of the ramping part of the waveform (see Sec. \ref{sec:exponential}). 
 (b) Softening shear response i.e $\tau''(|\gamma|)<0$. Figure is a representative one for a general softening solid. The dashed lines represent the response of a material with a linear shear stress response. }
\label{fig:spreading_shock}
\end{center}
\end{figure}

\noindent The shear wavespeed $c$  is a function of the local shear strain $\gamma(X_2,t)$. Hence, if the local wavespeed increases along the wave propagation direction, a given wave form will spatially spread over time. 
On the other hand, if the wavespeed decreases along the propagation direction, the waveform would steepen spatially over time and can eventually develop a local discontinuity when a shock forms. Accordingly, for a shear wave travelling in the positive $X_2$ direction, shocks can form when $\pdv{c}{X_2} =\dv{c}{\gamma}\pdv{\gamma}{X_2} < 0$. Hence, using 
the expression for $c(\gamma)$ from \eqref{waveeqn} and the fact that $\tau(\gamma)$ is an odd function, we obtain the condition for shear shock formation as\footnote{Here we have used the definitions $\tau ''(|x|) \equiv \eval{\tau''(\gamma)}_{\gamma = |x|}$, $\frac{\partial|\gamma|}{\partial X_2} \equiv \frac{\partial(\gamma \textrm{sign}(\gamma))}{\partial X_2}$, and we consider situations in which shear strain does not change sign. Also, whenever we discuss sign of $\tau''(\abs{\gamma})$ we consider non-zero shear strains. }
\begin{equation}\label{cond}
 \tau ''(|\gamma|) \frac{\partial|\gamma|}{\partial X_2}< 0   
\end{equation}
Formation of a shock thus depends on the loading program - loading or unloading the material, as defined by the sign of  $\frac{\partial|\gamma|}{\partial X_2}$ and on the nonlinear shear response of the material through the sign of $\tau''(|\gamma|)$. 
  A material with shear stress response such that $\tau ''(|\gamma|)>0$ will thus evolve a smooth shear loading waveform into a shock when it is being sheared, and one with a shear stress response such that  $\tau ''(|\gamma|)<0$ will produce a shear shock when unloaded from a sheared state.  
   Fig. \ref{fig:spreading_shock}  shows a loading shear waveform steepening into a shear shock versus spreading out, depending on the material constitutive shear response. If $\tau ''(\gamma) = 0 $, that is if the material shear response is linear ($\tau = \mu \gamma$) and the wavespeed is constant, the waveform will be relayed at the constant speed $c_s$ without steepening or spreading out  (dashed lines in Fig. \ref{fig:spreading_shock}). 

The most commonly used neo-Hookean \citep{rivlin1948large} and Mooney-Rivlin  \citep{mooney1940theory} hyperelastic models both predict a linear shear stress response, in which case there would be no nonlinear shear wave evolution. Biological tissues such as the brain have a strain stiffening shear response \citep{mihai2017family,mihai2015comparison,pogoda2014compression, storm2005nonlinear} and unsurprisingly commonly used hyperelastic models for soft solids and biological tissues such as the Gent model \citep{gent1996new,horgan2015remarkable} and the Fung model \citep{fung1993biomechanics} predict stiffening shear response i.e $\tau''(|\gamma|)>0$ (see \cite{mihai2015comparison}). Thus soft tissues and other materials whose constitutive response is well captured by these models would allow for shock formation in shear loading. Hence, in this work, we focus our attention on shear loading (as opposed to unloading from a pre-sheared state). Note that the velocity $V$ and the strain $\gamma$ will be of opposite signs by virtue of their definition. Thus when the elastic half-space is sheared along  positive $X_1$ direction (i.e $V>0$) it  develops a negative shear strain.

 While \eqref{cond} gives us the condition under which a shear wave can nonlinearly evolve into a shock, it does not provide information about the length and time scales over which the nonlinear evolution of the smooth waveform into a shock happens, or how this evolution depends on the material nonlinearity and the waveform. To that end, we will analyze the transient evolution of shear waves up to shock formation. We will then apply the analysis to different constitutive stress responses.

\subsection{Method of Characteristics}
\noindent To investigate the transient evolution of shear waves we follow the formulation in \cite{chu1964finite} (results recapitulated here) which makes use of the method of characteristics, we first introduce the auxiliary function $Q(\gamma)$ and its derivatives
\begin{equation}\label{Q_def}
    Q(\gamma)=\int\displaylimits_0^\gamma c(\gamma){\textrm d} \gamma, \qquad \frac{\partial Q}{\partial t}=c(\gamma)\frac{\partial \gamma}{\partial t}, \qquad  \frac{\partial Q}{\partial X_2}=c(\gamma)\frac{\partial \gamma}{\partial X_2}
\end{equation}
The monotonicity of $Q$ ensures that a unique value corresponds to a given shear strain $\gamma$. Further, $Q(\gamma)$ is an odd function ($Q(-\gamma)=-Q(\gamma)$) as $c(\gamma)$ is an even function. Using the above definitions, we can reduce the system of equations \eqref{wavesystem} to two ordinary differential equations whose solution is 
\begin{subequations}
\begin{align}
v + Q = f(\beta)  \quad &\textrm{and } v - Q = g(\alpha)   \\
\textrm{where} \quad \dd{\beta} =0 \textrm{\ \ along \ \ } \dv{X_2}{t} = -c \quad &\textrm{and } \dd{\alpha} =0 \textrm{\ \ along \ \ } \dv{X_2}{t} = c
\end{align} \label{sol}
\end{subequations}
The variables $\alpha$ and $\beta$ identify sets of characteristic curves described by $\dv{X_2}{t} = \pm c$ along which characteristic relations $\dd{v} = \pm c\dd{\gamma}$ hold. The general solution \eqref{sol} can be applied to our boundary value problem to solve for the transient shear wave evolution up to formation of shocks.  Upon formation of discontinuities the assumption of smoothness used to localize the integral form of momentum balance \eqref{momentum_conserv_integ_Lagrangian} breaks down, and thus the method of characteristics solution will not hold in regions with shocks.

\section{Boundary Value Problem}
\label{sec:bvp}
\noindent  Summarising our boundary value problem we have the initial and boundary conditions
\begin{equation}\label{Vt}
 v(X_2\ge0,t=0)=0 \quad , \quad \gamma(X_2\ge0,t=0)=0  \quad , \quad  v(X_2=0,t>0)=V(t)
 \end{equation} 
where we consider $V(t)$ to be a continuous piece-wise differentiable function of $t$. Using the method of characteristics, the solution for the velocity and strain fields  is\footnote{ There is no dependence on $\beta$ (from \eqref{sol}) as the initial shear strain in the body is constant (and zero).} \citep{chu1964finite}, 
\begin{subequations}
\begin{align}
\hat{v}(\alpha) &= V(\alpha) \label{vsol}\\
\hat{Q}(\alpha) &= - V(\alpha) \quad \xRightarrow{\eqref{Q_def}^{1}} \quad  V(\alpha)=-\int\displaylimits_{0}^{\hat\gamma(\alpha)} c(\gamma){\textrm d} \gamma \label{Qsol}
\end{align} \label{bothsol}
\end{subequations}
where fields written as a function of $\alpha$ have $\hat{()}$ accents and, as such, $\hat{Q}(\alpha) = Q(\hat{\gamma}(\alpha))$. According to \eqref{bothsol}, the velocity $v$ and shear strain $\gamma$ (or equivalently $Q$) are constant along characteristic lines identified by the characteristic variable $\alpha$.  The equation of a characteristic line identified by $\alpha$ is  \citep{chu1964finite}
\begin{equation}
X_2 = \hat{c}(\alpha)(t-\alpha) \label{charline}
\end{equation}
where $\hat{c}(\alpha) = c(\hat{\gamma}(\alpha))$ and the characteristics are labelled such that $\alpha = t$ on $X_2=0$ so that the characteristic $\alpha$ sets out from the loading surface at $t=\alpha$.
 \begin{figure}
\includegraphics[width=\textwidth]{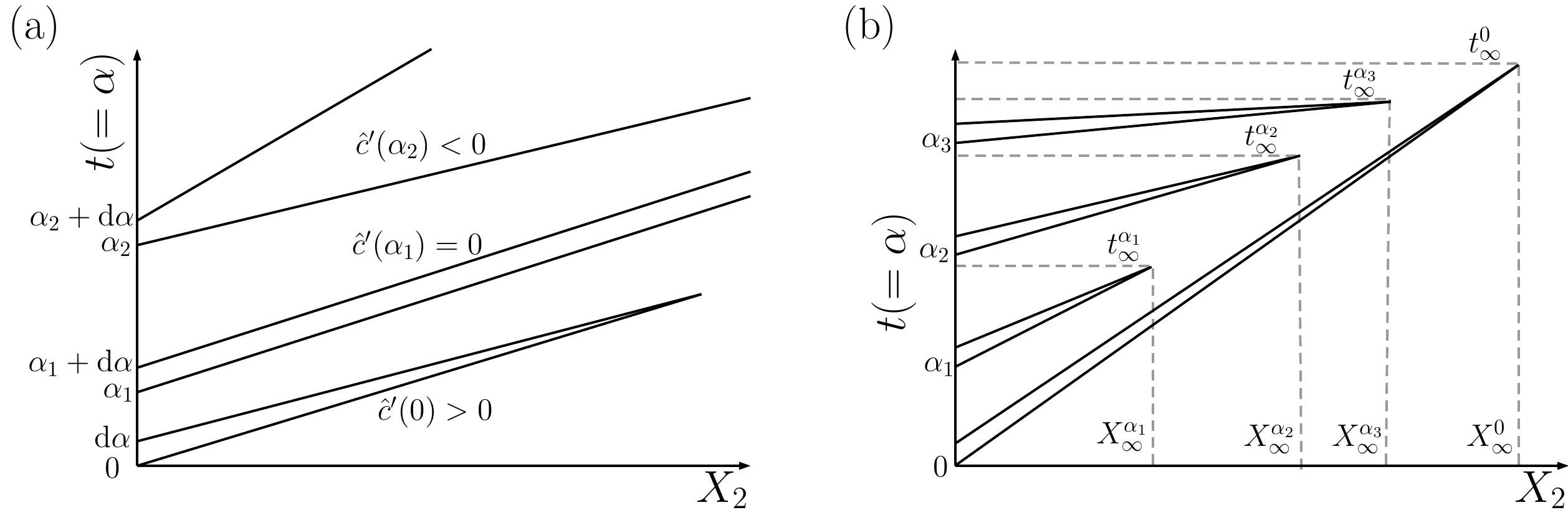}
\caption{(a) Characteristics with slope $1/c$ on the $t-X_2$ plane, described by \eqref{charline}, along which shear velocity and strain are constant. In regions where the characteristics run away from each other ($\hat{c}'<0$) the waveform  spatially spreads with time and in regions where their separation reduces ($\hat{c}'>0$) the waveform steepens. In regions where characteristics are parallel ($\hat{c}'= 0$) the waveform is locally relayed with no nonlinear evolution. When characteristics run into each other, shear shocks are formed, causing discontinuous strain and velocity fields. (b) The earliest time at which characteristics run into each other is the time at which the first shear shock is formed, i.e $t^{\textrm{shock}} = t_{\infty}^{\alpha_1}$ for the figure shown.  At later times ($t>t^{\textrm{shock}}$) the solution using method of characteristics breaks down and jump conditions would have to be applied at the location of the shock. The location at which the characteristics meet earliest can also be shown to be the shortest distance from the loading surface at which characteristics meet.}
\label{fig:characteristics}
\end{figure}
Equation \eqref{charline} determines a relation
\begin{equation}
\alpha = \alpha(X_2,t)
\end{equation}
which allows us to convert the solution field from functions of $\alpha$ to functions of $(X_2,t)$. 
When characteristics meet, a shock is formed, $\alpha$ becomes multivalued at a given $(X_2,t)$, 
and the solution \eqref{bothsol} is invalid beyond this time as previously discussed.  
 In this work we are only concerned with the transient nonlinear evolution of shear waves up to the first formation of a shock.

\subsection{Shock formation}
\label{sec:shock_formation}
\noindent We now apply the formulas from the previous section to  evaluate the  length and time scales needed for the development of a shock discontinuity. Representative characteristic lines in the $t-X_2$ plane are shown in Fig. \ref{fig:characteristics}(a). The slope of a characteristic line setting out at the loading surface $X_2=0$ at time $t=\alpha$ is $1/\hat{c}(\alpha)$. Consider two characteristics  described by some $\alpha>0$ and $\alpha+\dd{\alpha}>0$, that set out from the loading surface within an infinitesimal time separation $\dd{\alpha}$, with slopes $1/\hat{c}$ and $1/(\hat{c}+\dd{\hat c})$ respectively; if $\hat{c}' > 0$ they can meet and form a shock at a time
\begin{equation}
t^{\alpha}_{\infty} = \alpha + \frac{\hat{c}}{\hat{c}'} \ ,\qquad t^{\alpha}_{\infty} > \alpha \label{charmeettime}
\end{equation}
and at the location (using \eqref{charline})
\begin{equation}
X^{\alpha}_{\infty} = \hat{c}(t^{\alpha}_{\infty} - \alpha) = \frac{\hat{c}^2}{\hat{c}'} \label{charmeetdistance}
\end{equation}
where, from here on, quantities with the superscript $\alpha$ are assosciated with the characteristic line $\alpha$. It can be verified that the shock condition $\hat{c}'>0$ is consistent with the discussion  in Sec. \ref{subsec:shock_cond} (see \ref{app:shockcond}).

From \eqref{charmeettime} and \eqref{charmeetdistance}, the time and distance taken for characteristics to meet  depend inversely on $\hat{c}' = \dv{c}{\gamma}\hat{\gamma}'$, where $\dv{c}{\gamma}$ is smaller in magnitude for materials with weaker nonlinearity of shear response, and $\dv{\hat{\gamma}}{\alpha} = {\dv{\gamma(0,t)}{t}}$ is the strain rate generated at the loading surface which can be related to the velocity loading rate by differentiating \eqref{Qsol},
\begin{equation}
V'(\alpha) = -c\dv{\hat{\gamma}}{\alpha} \qquad \textrm{i.e} \qquad {\dv{V}{t}} = -c \dv{\gamma(0,t)}{t} \label{Vtogammarate}
\end{equation}
Accordingly, \eqref{charmeettime} and \eqref{charmeetdistance} can be written as
\begin{subequations}
\begin{align}
t^{\alpha}_{\infty} &=  \alpha + \frac{\hat{\psi}}{V'(\alpha)}  \quad \textrm{where} \quad \hat{\psi} = \psi({\hat{\gamma}}) \quad \textrm{and} \quad \psi(\gamma) = -{c^2}\bigg({\dv{c}{\gamma}}\bigg)^{-1} \label{charmeettime2}\\
X^{\alpha}_{\infty} &=  \frac{\hat{\psi}\hat{c}}{V'(\alpha)} \label{charmeetdistance2}   
\end{align} \label{meet_group}
\end{subequations}
Functions of $\alpha$ can be rewritten as functions of the strain and vice versa using \eqref{Qsol}, i.e $\hat{F}(\alpha) = F(\hat{\gamma}(\alpha))$. Thus higher the loading rate ($|V'|$) and/or stronger the material nonlinearity (higher $|\dv{c}{\gamma}|$), the quicker the characteristics meet and over shorter distances.  

The velocity gradient on a given characteristic $\alpha$, denoted by $v^\alpha_{X_2}$, can be found using \eqref{vsol} and \eqref{charline} as function of the time $t(>\alpha)$ or location $X_2$,
\begin{equation}
v^\alpha_{X_2} = \pdv{\hat{v}(\alpha(X_2,t))}{X_2} = \hat{v}'\pdv{\alpha}{X_2} \quad , \quad v^\alpha_{X_2}(X_2)  = \frac{V'(\alpha)}{\hat c' X_2/\hat{c} - \hat{c}} \quad , \quad \tilde{v}^\alpha_{X_2}(t)= \frac{V'(\alpha)}{\hat c'(t-\alpha) - \hat{c}} \label{vel_grad_char}
\end{equation}
Similarly, the strain gradient, $\gamma^\alpha_{X_2}$, can be found using \eqref{Vtogammarate} and \eqref{vel_grad_char}, as
\begin{equation}
\gamma^\alpha_{X_2} = \pdv{\hat{\gamma}(\alpha((X_2,t))}{X_2} = \hat{\gamma}'\pdv{\alpha}{X_2}  = -\frac{V'(\alpha)}{\hat{c}}\pdv{\alpha}{X_2} = - \frac{v^\alpha_{X_2}} {\hat{c}} \label{strain_grad_char}
\end{equation}
Thus, using \eqref{charmeettime} or \eqref{charmeetdistance} in \eqref{vel_grad_char} and  \eqref{strain_grad_char}, it can be seen that when characteristics meet the magnitude of local spatial gradients of the velocity and strain become infinite.  The local acceleration on a given characteristic $\alpha$, denoted by $a^\alpha$, can be also written as a function of $X_2$ or $t$ using \eqref{vsol} and \eqref{charline},  
\begin{equation}
a^\alpha = \pdv{\hat{v}(\alpha(X_2,t))}{t} = \hat{v}'\pdv{\alpha}{t} \quad,\quad a^\alpha(X_2)  = \frac{V'(\alpha)}{1-(\hat c'/\hat c^2)X_2} \quad,\quad \tilde{a}^\alpha(t)= \frac{V'(\alpha)}{1-(\hat c'/\hat{c})(t-\alpha)} \label{acc_on_alpha}
\end{equation}
and its magnitude is also seen to become infinite when characteristics meet (using  \eqref{charmeettime} or \eqref{charmeetdistance} in \eqref{acc_on_alpha}).

The earliest time at which characteristics meet is when the first shock forms and the method of characteristics solution will not hold at later times. Fig. \ref{fig:characteristics}(b) shows a set of representative charactersitics starting out at different times $\alpha_i$ meeting at different times $t_\infty^{\alpha_i}$, the earliest time at which characteristics meet, $t_\infty^{\alpha_1}$, is the time at which the first shock forms. Thus, in the general case, the time at which the first shock is formed is given by 
\begin{equation}
t^{\textrm{shock}} =\min_{\alpha \in [0,\infty) }t^{\alpha}_{\infty}  \label{tshock}
\end{equation}
Using \eqref{charmeetdistance2}, the location at which it forms, denoted by $X_{\textrm{shock}}$, can be found as
\begin{equation}
X_{\textrm{shock}} =  \eval{X^{\alpha}_{\infty}}_{\alpha_{\min}^\infty} = \eval{\frac{\hat{\psi}\hat{c}}{V'(\alpha)}}_{\alpha_{\min}^\infty}
\textrm{where \ }\quad  \alpha_{\min}^\infty = \argmin_{\alpha \in [0,\infty) }t^{\alpha}_{\infty} \label{Xshock}
\end{equation}
The characteristics that start out at $t= \alpha_{\min}^\infty$ are the ones that meet earliest in time to form a shock. When $X^{\alpha}_{\infty}$ and $t^{\alpha}_{\infty}$ are smooth functions of $\alpha$, which is true when the material response and loading are smooth (from \eqref{meet_group}), we can differentiate \eqref{charmeetdistance} and use \eqref{charmeettime} to write
\begin{equation}
\dv{X^\alpha_\infty}{\alpha} = \hat{c}'(t^\alpha_\infty - \alpha) + \hat{c}\bigg(\dv{t^\alpha_\infty}{\alpha}  - 1\bigg) = \hat{c}\dv{t^\alpha_\infty}{\alpha}
\end{equation}
Thus the sign of  $\dv{X^\alpha_\infty}{\alpha}$ and $\dv{t^\alpha_\infty}{\alpha}$ will be the same and $\alpha_{\min}^\infty$ will also be a minimizer of $X^\alpha_\infty$, that is,
\begin{equation}
\alpha_{\min}^\infty = \argmin_{\alpha \in [0,\infty) }X^{\alpha}_{\infty}
\end{equation}
Hence, among the locations of intersection of characteristics, the location at which the first shock forms, $X_{\textrm{shock}}$, is also at the shortest distance from the loading surface, when the loading and material behaviour are smooth.

\subsection{Acceleration magnification}
\label{sec:acc_mag}
\noindent While the spatial steepening of a waveform ultimately results in a shock, we would also like to quantify the nonlinear evolution of the shear wave as it progresses into the material. One way to do this is to analyse the magnification in acceleration along a characteristic with respect to the initial imposed acceleration. Physical systems are finite in length and we might be interested in knowing if a shock can form within a certain length for a given loading and if not what the maximum magnification in acceleration or field gradients would be, since higher velocity and strain gradients or equivalently higher accelerations, can cause damage in the material. Also, real materials are not perfectly elastic, they have some viscosity which has a spatial smoothening influence on the evolving waveform, hence a pure shock (discontinous fields) is never achieved, so it makes sense to talk in terms of acceleration/field gradient magnification which is a staple of nonlinear wave evolution irrespective of whether viscosity is included in the modelling. 

The magnification in acceleration along a characteristic $\alpha$ with respect to the imposed acceleration at the loading surface ($a(X_2 = 0,t = \alpha)$), can be written as a function of progressing $t$ or $X_2$ using \eqref{acc_on_alpha},
\begin{equation}
    M^\alpha(X_2) =\frac{a^\alpha(X_2)}{a^\alpha(0)}=\frac{1}{1-(\hat c'/\hat c^2)X_2} \quad , \quad \tilde{M}^\alpha(t) =\frac{\tilde{a}^\alpha(t)}{\tilde{a}^\alpha(\alpha)} = \frac{1}{1-(\hat c'/\hat{c})(t-\alpha)} \label{mag_char}
\end{equation}
Using \eqref{vel_grad_char}, \eqref{strain_grad_char}, and \eqref{mag_char}, it can be shown that the magnification in the velocity and strain gradients along a characteristic $\alpha$, with respect to their values at the loading surface, is equal to the magnification in acceleration, $M^\alpha$,
\begin{equation}
\frac{v^\alpha_{X_2}(X_2)}{v^\alpha_{X_2}(0)}=\frac{1}{1-(\hat c'/\hat c^2)X_2} = M^\alpha \quad, \quad \frac{\gamma^\alpha_{X_2}(X_2)}{\gamma^\alpha_{X_2}(0)}= \frac{v^\alpha_{X_2}(X_2)}{v^\alpha_{X_2}(0)} = M^\alpha\label{mag_char2}
\end{equation}
Thus, henceforth, whenever we talk about magnification in acceleration, it is tantamount to talking about magnification in the velocity or strain gradients.

 From \eqref{mag_char}, 
 it can be seen that when $\hat{c}'>0$ (converging characteristics in the $t-X_2$ plane as shown in Fig. \ref{fig:characteristics}(a)), the magnitudes of velocity gradient, strain gradient, and acceleration, along a characteristic $\alpha$, magnify as the wave progresses into the material 
  and become infinite when characteristics meet and form a shock (seen by using \eqref{charmeettime} and \eqref{charmeetdistance} in \eqref{mag_char}). On the other hand, when $\hat{c}'<0$ (diverging characteristics in Fig. \ref{fig:characteristics}(a)), the magnitudes of acceleration and field gradient along the characteristic decrease, and the wave locally spreads out. If $\hat{c}'=0$ (parallel characteristics in  Fig. \ref{fig:characteristics}(a)) there is no magnification of fields. For a material with linear shear response $\hat{c}' = \dv{c}{\gamma}\hat{\gamma}' = 0$ everywhere and the imposed waveform is relayed with no nonlinear evolution at the constant linear elastic shear wavespeed ($v(X_2,t) = V(t - X_2/c_s)$ using \eqref{vsol} and \eqref{charline}). 
  Since we are considering the loading of strain stiffening materials in this paper, the discussion hereon is for\footnote{Recall that $\hat{c}'= \dv{c}{\gamma}\hat{\gamma}'$. At zero shear strain,  $\dv{c}{\gamma}$ can be zero, and thus $\hat{c}'$ can be zero.} $\hat{c}' \ge 0$.

From \eqref{mag_char}, the time at which a given magnification in acceleration, $M$, is achieved along a characteristic $\alpha$, denoted by $t^\alpha_{M}$ , is given by  
\begin{equation}
t^\alpha_{M} = \alpha + \lambda_M \frac{\hat{c}}{\hat{c}'} \qquad \textrm{where} \qquad \lambda_M = 1 - \frac{1}{M} \label{talphaM}
\end{equation}
and the location at which it is achieved is given using \eqref{charline} and \eqref{talphaM} as
\begin{equation}
X^\alpha_{M} = \hat{c} (t^\alpha_{M}  -\alpha) = \lambda_M \frac{\hat{c}^2}{\hat{c}'}  \label{XalphaM}
\end{equation}
Note that for $\hat{c}' \ge 0$ and $t<t^\alpha_\infty$, $M\in[1,\infty)$ using \eqref{mag_char} and thus $\lambda_M\in[0,1)$. The earliest time at which an acceleration magnification $M$ is realized in the material, $t_{M}$, and the characteristic along which it is realized earliest, $\alpha_{\min}^M$ , are given by
\begin{equation}
t_{M} = \min_{\alpha \in [0,\infty) }t^\alpha_{M}  \quad , \quad  \alpha_{\min}^M = \argmin_{\alpha \in [0,\infty) } t^\alpha_{M} \label{tm}
\end{equation}
and the location at which the acceleration magnification $M$ is realized earliest, $X_{M}$ , is given by (using \eqref{XalphaM} and \eqref{tm})
\begin{equation}
X_{M} = \eval{X^\alpha_{M}}_{\alpha_{\min}^M} = \lambda_M\eval{\frac{\hat{c}^2}{\hat{c}'}}_{\alpha_{\min}^M} \label{Xm}
\end{equation}
Using \eqref{Vtogammarate} we can rewrite \crefrange{talphaM}{Xm} as
\begin{subequations}
\begin{align}
t^{\alpha}_{M} &= \alpha + \lambda_M \frac{\hat{\psi}}{V'(\alpha)}  \quad , \quad t_{M} = \min_{\alpha \in [0,\infty) } t^{\alpha}_{M} \label{tm1} \\
X_{M} &= \lambda_M \eval{\frac{\hat{\psi}\hat{c}}{V'(\alpha)}}_{\alpha_{\min}^M} \label{Xm2}
\end{align} \label{mag_soln}
\end{subequations} 
\noindent When the loading and material behaviour are smooth, we can differentiate \eqref{XalphaM} and use \eqref{talphaM} to write
\begin{equation}
\dv{X^\alpha_M}{\alpha} = \hat{c}'(t^\alpha_M - \alpha) + \hat{c}\bigg(\dv{t^\alpha_M}{\alpha}  - 1\bigg) = \hat{c}\bigg(\dv{t^\alpha_M}{\alpha} + (\lambda_M -1) \bigg) \label{dXalphaM_todtalphaM}
\end{equation}
Note that for a finite magnification $M$, we have $\lambda_M < 1$ and thus, the signs of $\dv{X^\alpha_M}{\alpha}$ and $\dv{t^\alpha_M}{\alpha}$ need not be the same. Hence, in general, $\alpha^M_{\min}$ need not be the minimizer of $X^\alpha_M$ and thus, the location at which a given finite acceleration magnification is realized earliest in time need not be the shortest distance from the loading surface at which it is realized.
The analysis for shock formation from the previous section is related to the acceleration magnification analysis by taking the limit $M\to \infty$, namely,
\begin{equation}
\{t^{\alpha}_{\infty} , \ t ^{\textrm{shock}}, \  {\alpha_{\min}^\infty},\  X_\textrm{shock} \} = \lim_{M\to\infty} \{t^{\alpha}_{M},\ t_M, \  {\alpha_{\min}^M}, \  X_M  \} \label{Mtoinfty}
\end{equation}

From \crefrange{vel_grad_char}{acc_on_alpha} and \crefrange{mag_char}{mag_char2}, it can be seen that a local discontinuity in the imposed waveform ($V' \to \infty$) would mean infinite local acceleration and field gradients but not necessarily an infinite magnification. By basing our investigation on analysis of the acceleration magnification \eqref{mag_char}, we distinguish between two types of discontinuities: those that are evolved due to nonlinearity of the material; and those that are imposed by a discontinuity in the loading waveform $V(t)$. The discussion of shock waves in this paper refers to the former, which is associated with the intersection of characteristics, and in turn represents the limit of acceleration/field gradient magnification becoming infinite.
The question may arise as to why one might want to draw the distinction between an imposed and a nonlinearly evolved discontinuity; in a physical system with even minute viscosity a high imposed local gradient would be immediately smoothened and a linear material will not try to steepen the smooth wave into a shock, whereas in a nonlinear material (with $\hat{c}'<0$), the material would continuously try to steepen the waveform, increasing the strength of the field gradients/acceleration with time while the viscosity tries to dissipate them away \citep{bland1965shock}.

\subsection{Immediate shock formation}
\noindent If the very first characteristics ($\alpha \to 0^+$) meet immediately, then a shock is instantaneously formed at $(t,X_2)=(0,0)$ and there would be no need to perform the minimisation in \eqref{tshock} and \eqref{Xshock}. To see when this could happen we first find the time it takes for the very first characteristics to meet by setting $\alpha \to 0^+ $ in \eqref{charmeettime2}. Without loss of generality, we make the assumption $V\ge0, \gamma\le0$ henceforth. When the initial shear strain is zero, we have $\hat{\psi}(0) = \psi(0)$, yielding 
\begin{equation}
t^{0}_{\infty} = \lim\limits_{\alpha \to 0^+} \frac{\hat{{\psi}}(\alpha)}{V'(\alpha)}  = - \ c_s^2 \lim\limits_{\substack{\alpha \to 0^+ \\\gamma \to 0^- }}{\bigg( V'(\alpha) \dv{c}{\gamma}\bigg)^{-1}}\label{tshockzero}
\end{equation}
where we have used $\eqref{charmeettime2}^{2,3}$ and \eqref{c_s_first}.
Since  $\dv{c}{\gamma}$ is bounded near zero shear strain, the only way for a shock to immediately form is when the initial loading rate, $V'(0)$, is infinite. However, this condition alone is insufficient, since for common materials the slope $\dv{c}{\gamma} \to 0$ may balance the loading singularity, leading to $t^{0}_{\infty}>0$,  as seen by examining the above equation.
The distance at which the very first characteristics meet is given, using \eqref{charmeetdistance}, as 
\begin{equation}
X^{0}_{\infty} = c_s t^{0}_{\infty} \label{Xshockzero}
\end{equation}
The time taken for the very first characteristics to realize an acceleration magnification $M$, and the distance at which it is realized, are similarly given by setting $\alpha \to 0^+$ in \eqref{tm1} and \eqref{Xm2}
\begin{equation}
t^{0}_{M} = \lambda_M \lim\limits_{\alpha \to 0^+} \frac{\hat{{\psi}}(\alpha) }{V'(\alpha)}  = -{\ c_s^2 \lambda_M} \lim\limits_{\substack{\alpha \to 0^+ \\\gamma \to 0^- }}{\bigg( V'(\alpha) \dv{c}{\gamma}\bigg)^{-1}} \quad , \quad X^{0}_{M} = c_s t^{0}_{M} \label{tMzero} 
\end{equation}
It can be seen from \eqref{tshockzero} and \eqref{tMzero} that when the very first characteristics meet immediately ($t^{0}_{\infty}=0$), any acceleration magnification $M\ge1$ is also immediately realized ($t^{0}_{M} = t_M = 0$). 
 
\subsection{Solution for a general family of loadings}
\label{sec:shearing_soft}
 \begin{figure}
\includegraphics[scale=0.6]{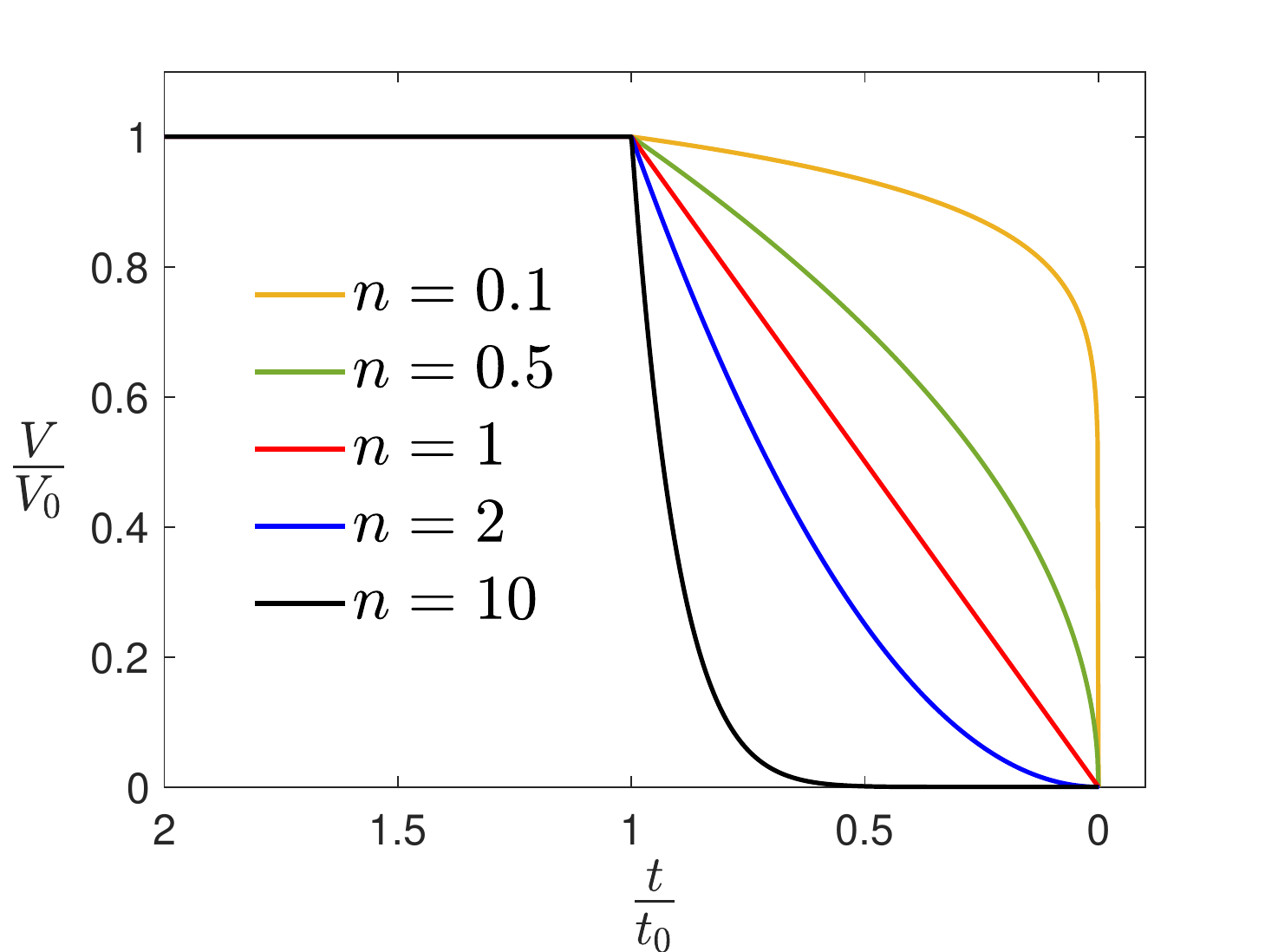}
\begin{centering}
\caption{Family of loading profiles described by \eqref{loadprofile}, $V_0$ is the ramp velocity, $t_0$ is the ramp time and $n$ is the loading exponent. The initial loading rate is infinite for $n<1$.}
\label{loadingprofiles}
\end{centering}
\end{figure}
\noindent Our goal is to study the effect of material behaviour and the loading waveform on the shock evolution. To that end, we choose a general loading profile
\begin{equation}
V(t)=V_0 \begin{cases}
 \left(\frac{t}{t_0}\right)^n & 0\leq t\leq t_0\ \\ \label{loadprofile}
1 & t\geq t_0  \end{cases}
\end{equation}
wherein the shearing velocity is ramped up in a continuous and monotonous fashion from zero at $t=0$ to a ramp velocity $V_0$ within a ramp time $t_0$, after which the velocity is held constant. Different values of the loading exponent $n>0$ reflect different ramping forms, as shown in Fig. \ref{loadingprofiles}, and in particular, $n=1$ corresponds to linear ramping where the acceleration is constant during the ramping phase. Note that as discussed earlier in Sec. \ref{sec:acc_mag}, an infinite initial loading rate ($t=0$, $n<1$) is not an evolved shock.

For the loading \eqref{loadprofile}, using \eqref{vsol} and \eqref{Qsol}, the velocity and strain fields can be seen to be constant for characteristics $\alpha>t_0$ and thus the characteristics are parallel 
and there is no magnification in acceleration along charactersitics $\alpha>t_0$  (setting $V'(\alpha) = 0$ in \eqref{tm1} gives $t^\alpha_{M} \to \infty$ for $t>t_0$ and $M>1$). Thus we can restrict the minimisation in \eqref{tm}  and \eqref{mag_soln} to the ramping phase
\begin{equation}
t_{M} = \min_{\alpha \in [0,t_0] } \alpha + \lambda_M \frac{\hat{\psi}}{V'(\alpha)}   \quad , \quad \alpha_{\min}^M = \argmin_{\alpha \in [0,t_0] } t^\alpha_{M}\label{tm2}
\end{equation}
Henceforth, unless mentioned otherwise, we consider  $\alpha \in [0,t_0]$. The loading is smooth in the interval of interest and hence, as discussed in Sec. \ref{sec:shock_formation}, $X_\textrm{shock}$ is also the shortest distance from the loading surface where characteristics intersect (assuming smooth material response).
 Using \eqref{Qsol}, we can specialize \eqref{Xm2} for the loading \eqref{loadprofile} as (see \ref{app:firstmeet})
\begin{equation}
X_{M} = \frac{t_0 \lambda_M }{n V_0^\frac{1}{n}} \eval{\frac{\psi c}{(-Q(\gamma))^{\frac{n-1}{n}}}}_{\gamma_{\textrm{min}}^M} \quad , \quad \gamma_{\textrm{min}}^M = \hat{\gamma}(\alpha_{\min}^M) \label{specialised_Xm}
\end{equation}

The time $t_M^0$ at which the very first characteristics attain acceleration magnification $M$, from \eqref{tMzero}, can be specialised for the loading \eqref{loadprofile} as (see \ref{app:firstmeet})
\begin{equation}
t^{0}_{M} =   \frac{c_s^{1+\frac{1}{n}} \ t_0 \lambda_M}{ {n V_0^{1/n}} } \lim_{\gamma \to 0^-} -\bigg(\dv{c}{\gamma}\bigg)^{-1} (-\gamma)^{\frac{1-n}{n}}
\label{tMzero2}
\end{equation} 
If $p_0$ is the order of the Taylor expansion of $\dv{c}{\gamma}$ for $\gamma \to 0^-$, i.e if
\begin{equation}
\eval{\dv{c}{\gamma}}_{\gamma \to 0^-} \to -c_0 (-\gamma)^{p_0} \label{slopeorder}
\end{equation} 
where $c_0$ and $p_0$ are constants\footnote{$c_0>0$ for a shock to form and $p_0 \ge 0$ since physically in the linear elastic limit we expect $|\dv{c}{\gamma}|$ to be bounded.}, then using \eqref{slopeorder} in \eqref{tMzero2} yields 
\begin{equation}
t^{0}_{M} =   \frac{c_s^{1+\frac{1}{n}} \ t_0 \lambda_M}{ {n c_0 V_0^{1/n}} } {(-\gamma)^{\frac{1}{n}-1-p_0}} 
\label{tMzero3}
\end{equation}
We can define a critical loading exponent, denoted by $n_{c} (\le 1)$,
\begin{equation}
n_{c} = \frac{1}{1+p_0} \label{n_crit}
\end{equation}
so that  for $n<n_c$, we have $t^{0}_{M} = X_M = 0$ (using \eqref{tMzero3} and $\eqref{tMzero}^2$), i.e there exists a material nonlinearity dependent loading exponent below which a shock forms immediately irrespective of the ramp velocity and ramp time. For $n>n_c$, we have $t^{0}_{M}, X^{0}_{M} \to \infty$ (using \eqref{n_crit} in \eqref{tMzero3}, $\eqref{tMzero}^2$), that is the initial characteristics are parallel and there is no acceleration magnification along them (there will still be acceleration magnification and shock formation along later characteristics).  For $n= n_c$, the values of $t^{0}_{M}$  and $X^{0}_{M}$  are finite and given by 
\begin{equation}
t^{0}_{M} \eval{}_{n = n_c} =   \frac{c_s^{2+p_0} \ t_0 \lambda_M (1+p_0)}{ { c_0 V_0^{1+p_0}} } \quad, \quad X^{0}_{M} \eval{}_{n = n_c} =   \frac{c_s^{3+p_0} \ t_0 \lambda_M (1+p_0)}{ { c_0 V_0^{1+p_0}} } 
 \label{tm0_ncrit}
\end{equation}
Essentially from \eqref{tMzero}, $t^{0}_{M}$ and  $X^{0}_{M}$ are inversely proportional to the limit of the product ${V'(\alpha) \dv{c}{\gamma}}$ near zero strain and thus there is a competition between the loading rate and the material's ability to resist nonlinear evolution decided by $|\dv{c}{\gamma}|$ (the smaller it is the less nonlinear the shear response and the easier it is for the material to relay the waveform as a spatially smooth one). 
For $n<n_{c}$, the loading rate singularity immediately overpowers the material's ability to resist shock formation (resisting  characteristics intersecting / acceleration magnification limit from becoming infinite). The weaker the nonlinearity, the higher the $p_0$ (from \eqref{slopeorder}) and thus smaller the $n_c$ using \eqref{n_crit}, meaning more powerful/singular loadings ($n<n_c$) are required to immediately form shocks for materials with weaker nonlinearity. 

 An essential function required for the minimisation in \eqref{tm2} is the derivative $\dv{t^{\alpha}_M}{\alpha}$, which for the loading \eqref{loadprofile} can be written as below (see \ref{app:derivativetm} for derivation),
\begin{equation}
\dv{t^{\alpha}_M}{\alpha} = \hat{G}(\alpha) = 1+ \hat{R}\lambda_M + \frac{\hat{\psi}(n-1)\lambda_M}{\hat{Q}n} \quad,\quad {R}(\gamma) = -{\dv{\psi}{\gamma}}{c^{-1}} \label{dtmdalpha}
\end{equation}

 In the following sections we will apply the formulas developed here to investigate transient evolution of shear waves into shocks for  materials with different constitutive responses. The Gent \citep{gent1996new} and Ogden \citep{ogden1972large} models are widely used for modelling the response of soft solids. 
 We first use an exponential 
 stress model which is not derived from any hyperelastic strain energy, but is one that greatly simplifies the analysis, and different features of the results therein carry over to the Gent and two parameter Ogden hyperelastic models (for parameter values that predict stiffening) analyzed subsequently. 
  Finally we apply the results of our analysis to a physical problem of shear impact of the brain.

\section{Analysis for different nonlinear shear constitutive responses}
\label{sec:applied_constit}
\subsection{Exponential stress model}
\label{sec:exponential}
 \begin{figure}
\includegraphics[width=\textwidth]{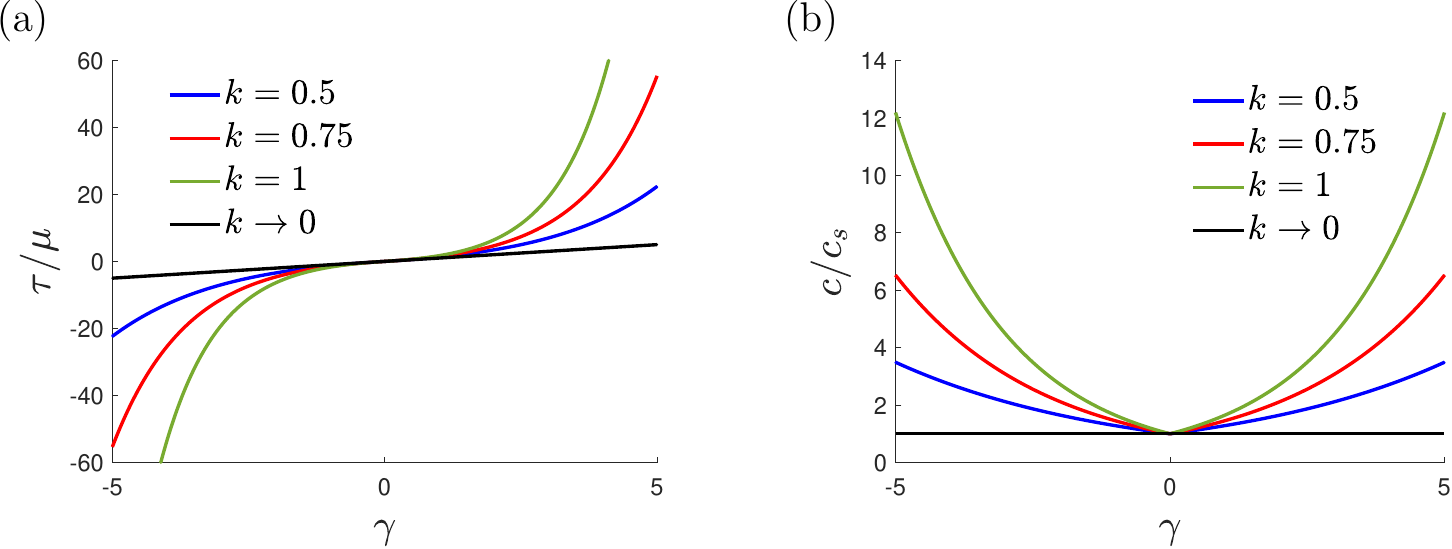}
\caption{Exponential stress model \eqref{exp_stress}:  Plots of (a) the dimensionless shear stress, and (b) the dimensionless shear wavespeed, as a function of the shear strain $\gamma$. The $k \to 0$ limit reduces to a Neo-Hookean shear response and $\gamma \to 0$ for any $k$ captures the linear elastic stress strain response.} \label{fig:exp_Stress}
\end{figure}
\noindent A simple function for the shear response that captures the nonlinear strain stiffening reported in experiments and allows for easy analytical tractability for our problem, is an exponential stress-strain response of the form
\begin{equation}
    \tau(\gamma) = \frac{\mu }{k}(e^{k\abs{\gamma}}-1)\textrm{sign}(\gamma) \label{exp_stress}
\end{equation}
where 
$k>0$ is a dimensionless parameter that quantifies the nonlinear stiffening. This function reduces to a linear elastic response in the limit of $\gamma \to 0$. However, it is not derived from any established hyperelastic strain energy function and is simply being used as a phenomenological model for analytical tractability; many features of the results carry over to the analysis for Gent and two parameter Ogden hyperelastic models in the subsequent subsections. For the response \eqref{exp_stress}, we have $\tau''(|\gamma|)>0$ and hence shocks will be formed in loading. 
Using \eqref{waveeqn}, the wavespeed is given by 
\begin{equation}
    c(\gamma) = c_s e^{\frac{k}{2}\abs{\gamma}} 
\end{equation}
Fig. \ref{fig:exp_Stress} shows plots of the stress response and the shear wavespeed for different values of the stiffening parameter $k$.

The derivative of the wavespeed is given by
\begin{equation}
 \dv{c}{\gamma} = \frac{c_s k }{2} e^{\frac{k}{2}\abs{\gamma}} \textrm{sign}(\gamma) \quad \textrm{for} \quad \gamma \ne 0  \label{c_slope_gent}
\end{equation}
Thus from \eqref{slopeorder}, we have $p_0 =0$ and $c_0 = c_s k/2$ for \eqref{c_slope_gent}, and using \eqref{n_crit} we get $n_c = 1$. Thus, for $n<1$, we have $t_{\textrm{shock}} = t_M = 0$ and $X_{\textrm{shock}}= X_M  =0$. This is not an artifact of the initial loading rate  being infinite for $n<1$ but instead represents the limit of acceleration magnification becoming infinite. In the subsequent sections for Gent and Ogden solids it will be shown that an infinite initial loading rate can still require a finite time and distance for the magnification in acceleration to increase or become infinite. Note that this deviation from conventional hyperelastic models essentially stems from the $\textrm{sign}(\gamma)$ factor in the stress response that allows for a quadratic term in the stress expansion whereas conventional energy expansions would only allow a cubic term. For $n\ge1$ we have to perform the minimization in $\eqref{tm2}^2$ to find $\alpha_{\textrm{min}}^M$.

Further, we write,
\begin{equation}
    Q(\gamma) = V_* (e^{\frac{k}{2}\abs{\gamma}}-1)\textrm{sign}(\gamma)  \quad \textrm{where} \quad V_* = \frac{2c_s}{k} \label{QandscaleV_exp}
\end{equation}
where we define a characteristic scaling velocity $V_*$ (not to be confused with velocity of a characterisitic).
As mentioned earlier we assume $V\ge0$, $\gamma\le0$. Using \eqref{Qsol}, we have
\begin{equation}
\gamma = -Q^{-1}(V(\alpha)) = -\frac{2}{k}\ln{\bigg(1+\frac{V}{V_*}\bigg)}  \label{gammaexp}
\end{equation}
Using \eqref{gammaexp}, we have for the exponential solid (refer \ref{app:exp})
\begin{equation}
\hat{c} = c_s \bigg(1+\frac{V}{V_*}\bigg) \quad , \quad  \hat{\psi}= V + V_* \quad , \quad \hat{R} =1 \label{exp_gamma_funcs}
\end{equation}
Combining eqs \eqref{dtmdalpha}, \eqref{exp_gamma_funcs} and \eqref{Qsol} we have
\begin{equation}
\dv{t^{\alpha}_M}{\alpha} = 1 + \lambda_M - \frac{(n-1)\lambda_M}{n}\bigg(1+\frac{V_*}{V}\bigg) \label{tderiv_exp}
\end{equation} 
For $n=n_c=1$ we have $\dv{t^{\alpha}_M}{\alpha}>0$ and thus $\alpha_{\textrm{min}}^M=0$.
Acceleration magnification and shock formation thus initiate at the leading edge of the waveform for $n=n_c$  and using \eqref{Xm} and  $\eqref{tm0_ncrit}^2$ we have
\begin{equation}
X_M = \eval{X_M^\alpha}_{\alpha^M_\textrm{min}} =  X^{0}_{M} \eval{}_{n = n_c} =   \frac{2c_s^{2} \ t_0  \lambda_M}{ { k V_0} } =  \frac{L_* \lambda_M}{ {m} } \label{expXmncrit}
\end{equation}
where we define the loading mach number $m$($\ge0$) and the characteristic length scale $L_*$ as 
\begin{equation}
 m = \frac{V_0}{V_*} \quad, \quad L_* = c_s t_0 \label{m_andL*}
\end{equation}
 Using \eqref{tderiv_exp} to perform the minimization in $\eqref{tm2}^2$ for $n>1$, we end up with (see \ref{app:exp})
  \begin{equation}
{\alpha_{\textrm{min}}^M} = \begin{cases}
            \begin{aligned}
t_0 \bigg(\cfrac{m_\textrm{th}}{m}\bigg)^{\frac{1}{n}}        \quad    & n > 1,\   m \ge m_{\textrm{th}}       \\ \label{expalphamins}
t_0 \phantom{\bigg(\cfrac{m_\textrm{th}}{m}\bigg)^{\frac{1}{n}} } & n > 1,\   0 \le m < m_{\textrm{th}}
            \end{aligned}
 \end{cases}
\end{equation}
where we have defined the the threshold mach number $m_{\textrm{th}}$ as 
\begin{equation}
 m_{\textrm{th}}  = \frac{(n-1)\lambda_M}{n+\lambda_M} \quad (n > 1) \quad , \quad   0 < m_\textrm{th} < \lambda_M (<1) \label{mthresh_exp}
\end{equation}
The threshold mach number is an increasing function of both the loading exponent $n$ and the target acceleration magnification $M (= 1 - 1/\lambda_M)$. Additionally, $m_{\textrm{th}}\to0$ as $n\to1$. 
From \eqref{expalphamins}, we see that for loadings with $n>1$, the earliest realization of target acceleration magnification or shocks can happen either at the the trailing edge of the waveform ($\alpha=t_0$) or in the ramping part ($0<\alpha<t_0$) depending on $m$ and $m_\textrm{th}$.

 Using \eqref{expalphamins} in \eqref{Xm2}, and the result \eqref{expXmncrit}, we have $X_M$, the location where an acceleration magnification of $M$ is realized earliest in the solid, non-dimensionalized by $L_*$ as (see \ref{app:exp})
 \begin{figure}
 \begin{centering}
\includegraphics[width=\textwidth]{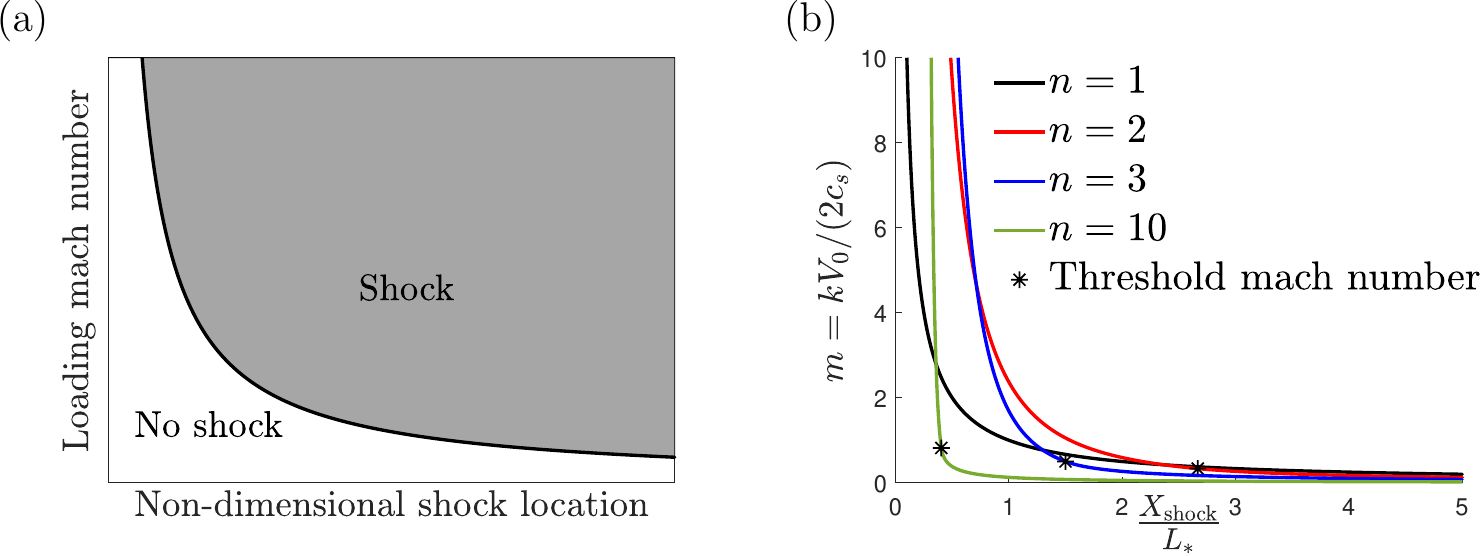}
\caption{(a) Representative non-dimensional phase map of the shock location for a given loading exponent $n\ge n_c$. (b) Non-dimensional maps of shock location as a function of the loading  mach number $m$, for various loading exponents, using the result \eqref{Xshockseps} for the exponential stress model \eqref{exp_stress}. For $n>1(n_c)$, a threshold mach number ($m_\textrm{th}$) separates two different functional forms of $X_\textrm{shock}/L_*$. In different loading mach number regimes different loading exponents might delay shock formation.} \label{fig:exp_Stress_Shock}
\end{centering}
\end{figure}
  \begin{equation}
\frac{X_M}{L_*} = \begin{cases} \label{Xmsexp}
\phantom{0}0  & n <1,\ m\ge0 \\[2pt]
  \cfrac{\lambda_M}{m} &   n = 1,\ m\ge0\\[5pt] 
 \cfrac{n(1+\lambda_M)^2}{(n-1)(n+\lambda_M)} \bigg(\cfrac{m_{\textrm{th}}}{m}\bigg)^{\frac{1}{n}} & n> 1,\ m \ge m_{\textrm{th}}\\[10pt] 
\cfrac{\lambda_M}{n}(2+m+ \frac{1}{m}) & n > 1,\ 0 \le m < m_{\textrm{th}}  
 \end{cases} 
\end{equation}
The solution is continuous at $n\to1^+$ and at $m \to m_{\textrm{th}}^-$ but is discontinuous at $n \to 1^-$ (see \ref{app:exp}).
 Setting $M\to\infty$, that is $\lambda_M=1$, in \eqref{Xmsexp} gives the non-dimensionalized shock location,
  \begin{equation}
\frac{X_{\textrm{shock}}}{L_*} = \begin{cases}
\phantom{;}0 & n <1,\ m\ge0 \\[2pt]
    \cfrac{1}{m} &   n = 1,\ m\ge0\\[5pt]  \label{Xshockseps}
 \cfrac{4n}{n^2-1} \bigg(\cfrac{m_{\textrm{th}}}{m}\bigg)^{\frac{1}{n}} & n> 1,\ m \ge m_{\textrm{th}}\\[10pt] 
\frac{1}{n}(2+m+ \frac{1}{m}) & n > 1,\ 0 \le m < m_{\textrm{th}}  
 \end{cases} 
\end{equation}
where $m_\textrm{th} =  (n-1)/(n+1)$. The expressions in \eqref{Xmsexp} and \eqref{Xshockseps} provide us the location of first realization of given acceleration magnification and of first shock formation respectively. 
\begin{figure}[!ht]
\begin{center}
\includegraphics[width=\textwidth]{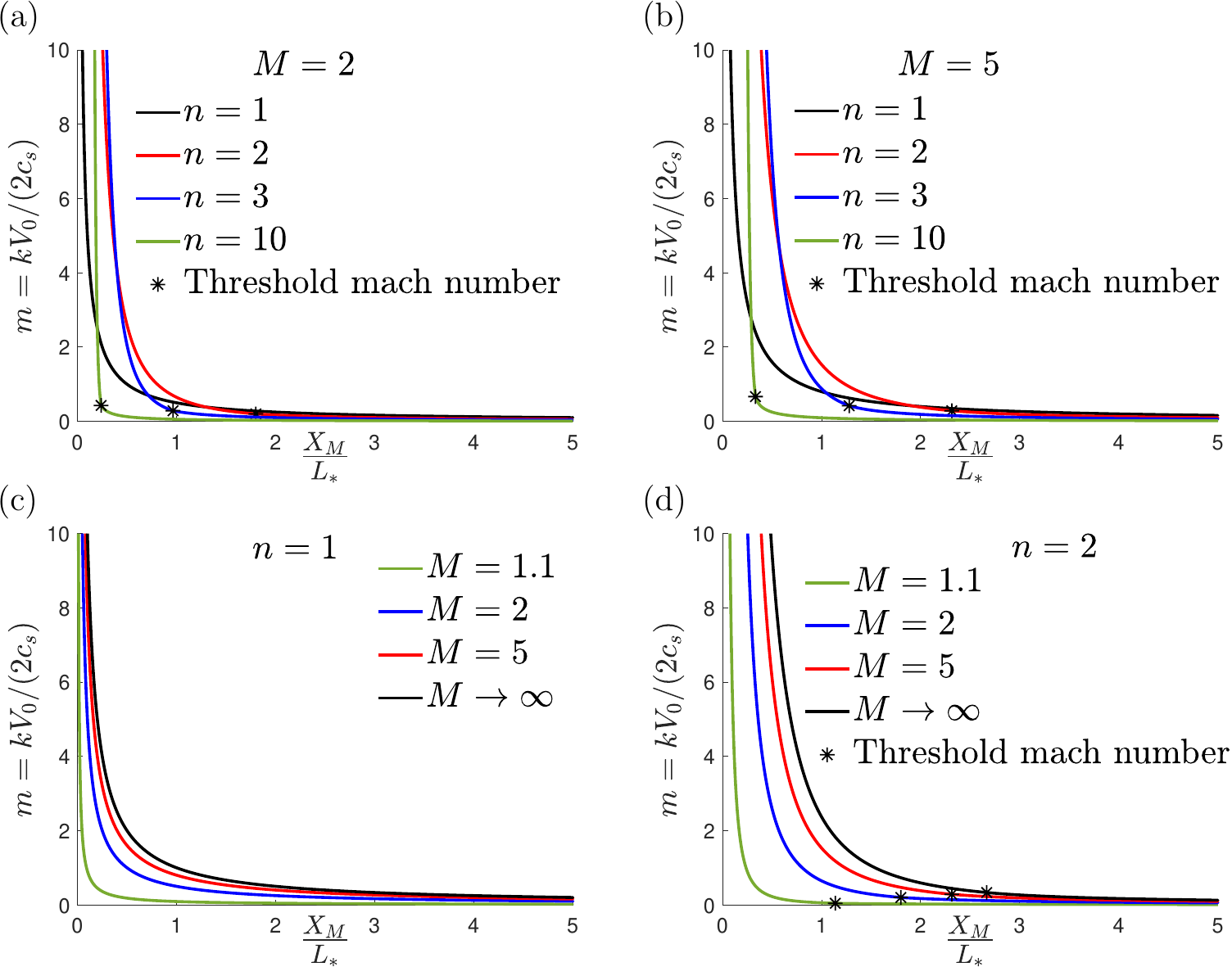}
\end{center}
\caption{Exponential stress model \eqref{exp_stress}: Maps of the dimensionless distance of first realization of acceleration magnification $M$, for varying loading mach number $m$, using \eqref{Xmsexp}. Maps for different loading exponents $n$ with (a) $M=2$, and (b) $M=5$. Maps at  different $M$ for (c) $n=1$($n_c$) and (d) $n=2$. In (c),(d), curves for $X_M$ quickly converge to that of the shock  location $(M\to\infty)$ at higher values of $M$. }
\label{fig:xshocks_exp}
\end{figure}

\textbf{Non-dimensional maps:}
The result in eq. \eqref{Xshockseps} allows us to make non-dimensional phase maps, such as the representative one shown in Fig. \ref{fig:exp_Stress_Shock}(a) for a given loading exponent  $n\ge n_c$, that would allow us to determine whether or not a shock would form within a given distance from the loading surface. The maps can serve as a guide in design problems.  For example, 
they can be used to estimate the maximum length one could design for, while avoiding shock formation for a given material and loading\footnote{Neglecting reflections in a finite dimensional system.}. Similarly they can be used to design loadings that will avoid shock formation given a material and a length scale, or to choose a material that will avoid shocks within a given length scale for a given loading. In Fig. \ref{fig:exp_Stress_Shock}(b), we use \eqref{Xshockseps} to make such maps for different values of the loading exponent $n$, 
where we have the loading mach number $m = kV_0/(2c_s)$ on the vertical axis and the distance for first shock formation non-dimensionalized by $L_*=c_st_0$ on the horizontal axis.  We can see that $X_\textrm{shock}/L_*$ reduces with increasing loading mach number for any $n$ (can be shown for any $n\ge1$ from \eqref{Xshockseps}). For $n>1(n_c)$ there is a threshold mach number separating different functional forms of $X_\textrm{shock}$. 

\textbf{Effect of material parameters:}
We first fix the loading waveform and study the effect of material parameters on shock evolution. The material response is quantified by the nonlinearity parameter $k$, which influences 
the loading mach number $m$,  and by the linear elastic shear modulus $\mu$, which appears through $c_s$ in both $m$ and $L_*$. Since $X_\textrm{shock}$ reduces with increasing $m$ and reducing $L_*$, higher material nonlinearity (higher $k$) and lower $\mu$ reduce the distance taken for first shock formation.  

\textbf{Effect of loading:} Next, we fix the material properties and analyze the effect of the loading parameters. The loading mach number $m$ increases with $V_0$ and hence, a higher ramp velocity reduces $X_\textrm{shock}$ (for a fixed $n$ and $t_0$).  The ramp time $t_0$ enters through $L_* = c_s t_0 $, and thus $X_{\textrm{shock}}$ scales linearly with $t_0$ for given $V_0$ and $n$. Thus as $t_0 \to 0$, we have $X_{\textrm{shock}} \to 0$ , i.e as the loading waveform approaches a discontinuity, a pure shock is immediately formed. 
The loading exponent $n$ seems to have the most profound and non-trivial effect on shock evolution. For example, it can be seen that going from $n\to 1^-$ to $n=1 (n_c)$ may be the difference between a shock immediately forming versus at a distance $L_*/m$ (see \eqref{Xshockseps}). From Fig. \ref{fig:exp_Stress_Shock}(b), it  can be seen that at a given loading mach number the trend in first shock location with respect to the loading exponent is non-monotonous. Also, the maps for different loading exponents intersect each other, meaning that there are different loading mach number regimes in which different loading exponents could delay shock formation. 
This could have implications for design of protective structures, if we know the mach regime we are operating in for a given problem, we can select a target waveform (by choosing $n$) that could delay the shock formation. 

\textbf{Acceleration magnification:}
Using \eqref{Xmsexp} we can make non-dimensional maps as shown in Fig.  \ref{fig:xshocks_exp}, where we have the loading mach number on the vertical axis and the location of first realization of a given acceleration magnification $M$, non-dimensionalized by $L_*$ on the horizontal axis. From Figs. \ref{fig:xshocks_exp}(a),(b), it can be seen that $X_M$ shows a similar qualitative dependence on the material and loading parameters as $X_\textrm{shock}$ and the commentary for shock formation carries over. However, when $n>1 (n_c)$ and $m>m_\textrm{th}$, the distance at which a given finite acceleration magnification is realized earliest in time need not be the shortest distance at which it is achieved (see \ref{app:exp}). In Figs. \ref{fig:xshocks_exp}(c),(d), we have maps of ${X_M}/{L_*}$ for two different loading exponents and a range of increasing $M$, the limit $M \to \infty$ corresponds to  ${X_{\textrm{shock}}}/{L_*}$. For a given material and loading it takes longer distances for first realization of higher acceleration magnifications, as one would expect. More importantly, it can be seen that the evolution from high acceleration magnifications ($M \sim 5$) to a shock wave ($M\to\infty$) happens over much shorter distances than the distance taken for realization of those high acceleration magnifications (starting from $M=1$). This is true for any $n\ge1 (n_c)$ though only two loading exponents have been shown here. If the ramping is linear ($n=1$) so that the acceleration imposed at the loading surface is constant during ramping, maps like Figs. \ref{fig:xshocks_exp}(c),(d) can also be used to predict the maximum acceleration realized within a length scale\footnote{Neglecting reflections in a finite dimensional system.}. This might be useful for application to biological systems where the damage tolerance might be quantified in terms of maximum acceleration.

\textbf{Post first shock formation:}
 Note that we have restricted our discussion so far to the onset of shock formation and have not touched on the strength of the shock. The first shock that forms would be weak, with the jump in velocity and strain fields being small. Assuming a steady state pure shock ultimately propagates into the material, the weak initial shock would eventually grow into a stronger shock with a jump in velocity from $0$ to $V_0$ across the shock, and this would happen over longer distances for materials with weaker nonlinearity of shear response. The strength of the steady state shock would be higher for higher ramp velocity $V_0$ and a consideration of the shock strength would be pertinent for materials with viscosity as it would be tougher to dissipate away larger discontinuities. Post the first shock formation we would have to apply the integral form of balance laws in a small volume surrounding the shock while using \eqref{sol} in the smooth regions, however  $f(\beta)$ (from \eqref{sol}) will no longer be constant behind the shock and there will be information travelling along negative $X_2$ direction as well (see \ref{app:postshock} for a relevant discussion). The problem becomes analytically intractable post shock initiation and would have to be studied numerically wherein numerical viscosity effects are inevitable.  

\subsection{Gent model}

    \begin{figure}
\includegraphics[width=\textwidth]{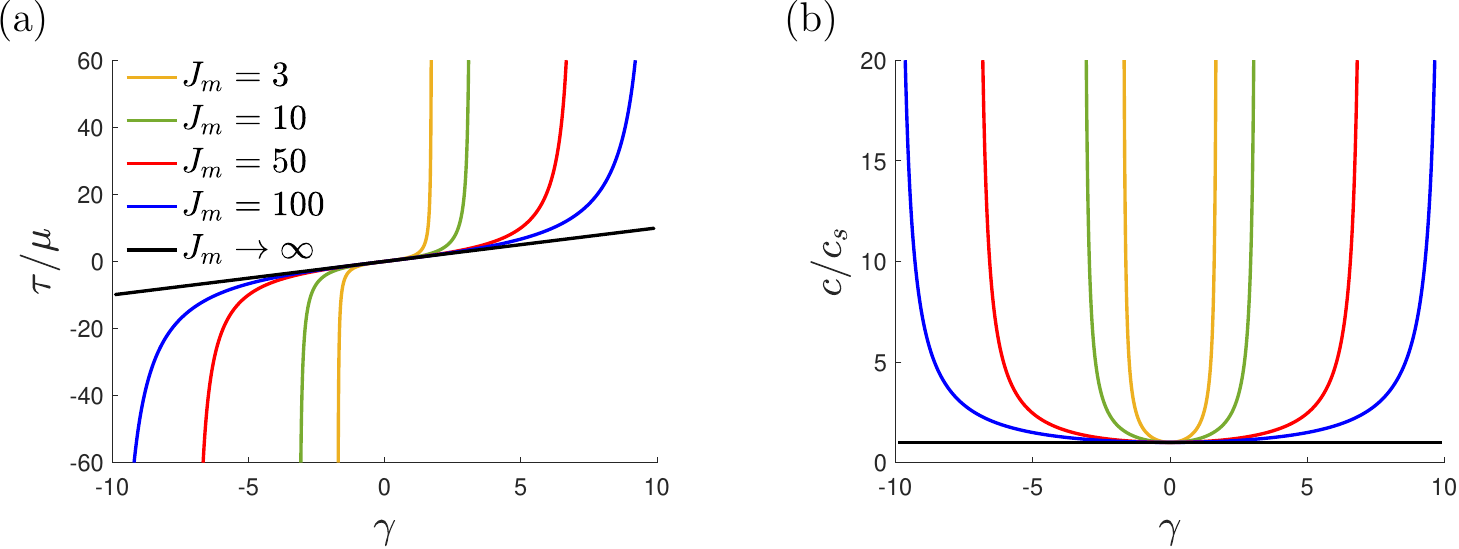}
\caption{Gent model \eqref{Gent_Stress}: Plots of (a) the dimensionless shear stress, and (b) the dimensionless shear wavespeed (same plot legend as for  (a)), as a function of the shear strain $\gamma$. 
 The stresses and wavespeeds become unbounded as the magnitude of shear strains approach the locking limit $\sqrt{J_m}$. Smaller the $J_m$, the more nonlinear the shear response of the solid. } 
\label{fig:Gentstress}
\end{figure} 
%
The popular and widely used Gent hyperelastic model \citep{gent1996new} is based on the concept of limiting chain extensibility where the strain energy density function is designed to have a singularity when the first invariant of $\nten B$, reaches a limiting value $I_m = J_m +3$,
\begin{equation}
W = -\cfrac{\mu J_m}{2} \ln\left(1 - \cfrac{I_1-3}{J_m}\right) \label{gent_W}
\end{equation}
For the simple shear deformation \eqref{defgradientshear}, this translates to limiting 
the magnitude of shear strain to a locking strain of $\gamma_m = \sqrt{J_m}$ as the shear stress keeps increasing (using \eqref{gent_W} in \eqref{Piola_incom}),
\begin{equation}
    \tau(\gamma) = \cfrac{\mu J_m \gamma}{J_m - \gamma^2} \quad\textrm{where}\quad J_m = \gamma_m^2 \label{Gent_Stress}
\end{equation}
Plots of the shear stress-strain curves for various values of $J_m$ are shown in Fig. \ref{fig:Gentstress}(a), the smaller the $J_m$ the more nonlinear the shear response. When $J_m \to \infty$, the strain energy density function in \eqref{gent_W} reduces to that of the neo-Hookean solid, yielding a linear shear stress-strain response. For finite $J_m$, we have $\tau''(|\gamma|)>0$ and hence, a solid whose shear response is well modelled by the Gent model will produce shear shocks in loading. We define a fractional shear strain $\delta$ which is the ratio of the shear strain and the locking strain $\gamma_m$,
\begin{equation}
    \delta  = \frac{\gamma}{\gamma_m} = \frac{\gamma}{\sqrt{J_m}} \quad ,\quad |\delta| < 1
\end{equation}
    \begin{figure}
    \begin{center}
\includegraphics[width=0.5\textwidth]{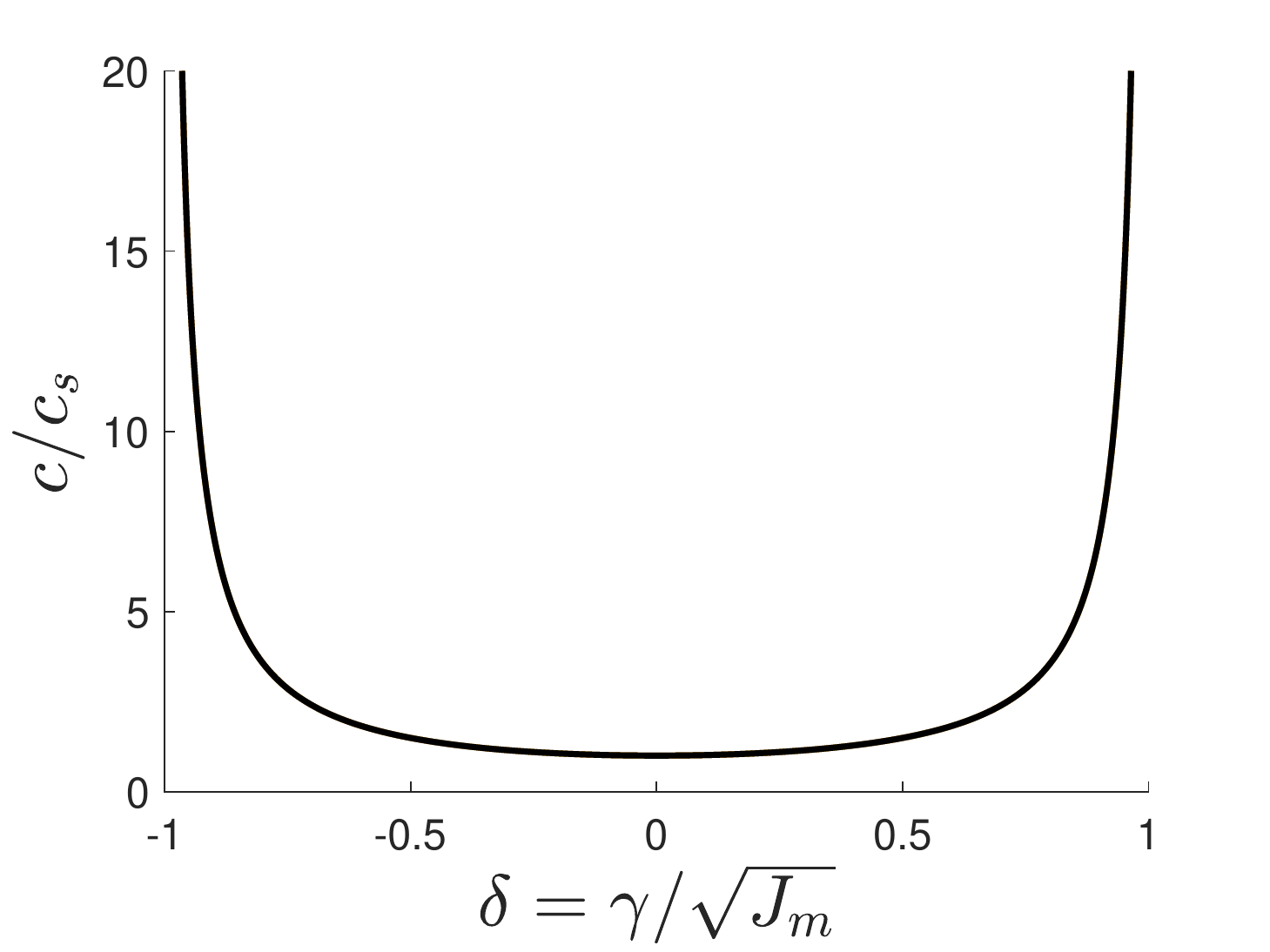}
\caption{Gent model \eqref{Gent_Stress}: Plot of the dimensionless wavespeed as a function of the fractional shear strain $\delta = \gamma/\sqrt{J_m}$, using \eqref{gent_c}.
}      \label{fig:Gent_c}
    \end{center}
\end{figure} 
Once again, without loss of generality we assume $V\ge0$ and thus $\gamma, \delta \le 0$.
Functions of the shear strain $f(\gamma)$ can be written as $\tilde{f}(\delta) = f(\delta \sqrt{J_m})$ where the tilde superscript denotes that the argument of the function is $\delta$. 
Using \eqref{waveeqn}, the wavespeed for the stress response \eqref{Gent_Stress} is given by 
\begin{equation}
    c(\gamma)  =V_* \cfrac{\sqrt{J_m + \gamma^2}}{J_m - \gamma^2}  \quad \textrm{where} \quad V_* = c_s \sqrt{J_m} \quad; \quad \tilde{c}(\delta) = c_s \cfrac{\sqrt{1 + \delta^2}}{1 - \delta^2} \label{gent_c}
\end{equation}
Here, as in \eqref{QandscaleV_exp}, we  have defined a stiffening parameter dependent characteristic scaling velocity $V_*$.
Plots of the wavespeed as a function of the shear strain are showin in Fig \ref{fig:Gentstress}(b) for various values of $J_m$. All of these curves collapse onto a single curve when plotted as a function of $\delta$ as shown in Fig. \ref{fig:Gent_c}. 
We can also write \eqref{Qsol} in terms of the fractional shear strain, 
\begin{equation}
\tilde{Q}(\delta) = - V(\alpha)   \label{Qsolgent}
\end{equation}
For the Gent stress response \eqref{Gent_Stress}, we have 
\begin{equation}
\tilde{Q}(\delta) = V_* \log\Bigg({\cfrac{ \bigg(\frac{\sqrt{1 + \delta^2} + \delta\sqrt{2}}{\sqrt{1 + \delta^2} - \delta\sqrt{2}}\bigg)^\frac{1}{\sqrt{2}}}{{\delta+\sqrt{1 + \delta^2}}}}\Bigg) \label{Qdeltagent} 
\end{equation}
where $\tilde{Q}$ is an odd function. We can divide \eqref{Qsolgent} and \eqref{Qdeltagent} by $V_*$ to write
\begin{equation}
\tilde{q}(\delta) = \frac{\tilde{Q}(\delta)}{V_*} =  - \frac{V}{V_*}  \quad \Rightarrow \quad \delta = -\tilde{q}^{-1}\bigg({\frac{V}{V_*}}\bigg) \quad \textrm{where} \quad \tilde{q}(\delta) = \log\Bigg({\cfrac{ \bigg(\frac{\sqrt{1 + \delta^2} + \delta\sqrt{2}}{\sqrt{1 + \delta^2} - \delta\sqrt{2}}\bigg)^\frac{1}{\sqrt{2}}}{{\delta+\sqrt{1 + \delta^2}}}}\Bigg) \label{qsolGent}
\end{equation} 
where $|\tilde{q}(\delta)|$ can be thought of as the local mach number dependent on the local velocity field $V(\alpha(X_2,t))$ (different from the loading mach number $m$ which depends on ramp velocity $V_0$ and is constant for a given loading). Equation \eqref{Qsolgent} provides a mapping $\delta = \hat{\delta}(\alpha)$, allowing us to switch function arguments between $\delta$ or $\alpha$ i.e $\hat{F}(\alpha) = \tilde F(\hat{\delta}(\alpha))$. Thus, we can rewrite \eqref{specialised_Xm} as
\begin{equation}
X_{M} = \frac{t_0 \lambda_M}{n V_0^\frac{1}{n}} \eval{\frac{\tilde{\psi}\tilde{c}}{(-\tilde{Q}(\delta))^{\frac{n-1}{n}}}}_{\delta_{\textrm{min}}^M} \quad \textrm{where} \quad \delta_{\min}^M = \hat{\delta}(\alpha_{\min}^M) \label{specialisedXmGent}
\end{equation}
which can be evaluated further for the Gent solid as (see \ref{app:gent})
\begin{equation}
X_{M} = \cfrac{L_*\lambda_M}{n m^{\frac{1}{n}}} \eval{\cfrac{K(\delta)}{(-q(\delta))^{\frac{n-1}{n}}}}_{\delta_{\textrm{min}}^M} \quad, \quad K(\delta) = {\cfrac{-(1+\delta^2)^2}{\delta(3+\delta^2)(1-\delta^2)}}   \label{specialisedXmGent2}
\end{equation}
The loading mach number $m$ and characteristic length scale $L_*$ are defined in \eqref{m_andL*}.

The derivative of the wavespeed \eqref{gent_c} is given by
\begin{equation}
 \dv{c(\gamma)}{\gamma} = \cfrac{V_*\gamma(3J_m+\gamma^2)}{(J_m - \gamma^2)^2\sqrt{J_m+\gamma^2}} \quad,\quad \dv{c(\gamma)}{\gamma}  \to \cfrac{3 c_s \gamma}{J_m} \quad {\textrm{as}} \quad \gamma \to 0^- \label{gentdcdg}
\end{equation}
\begin{figure}
\includegraphics[width=\textwidth]{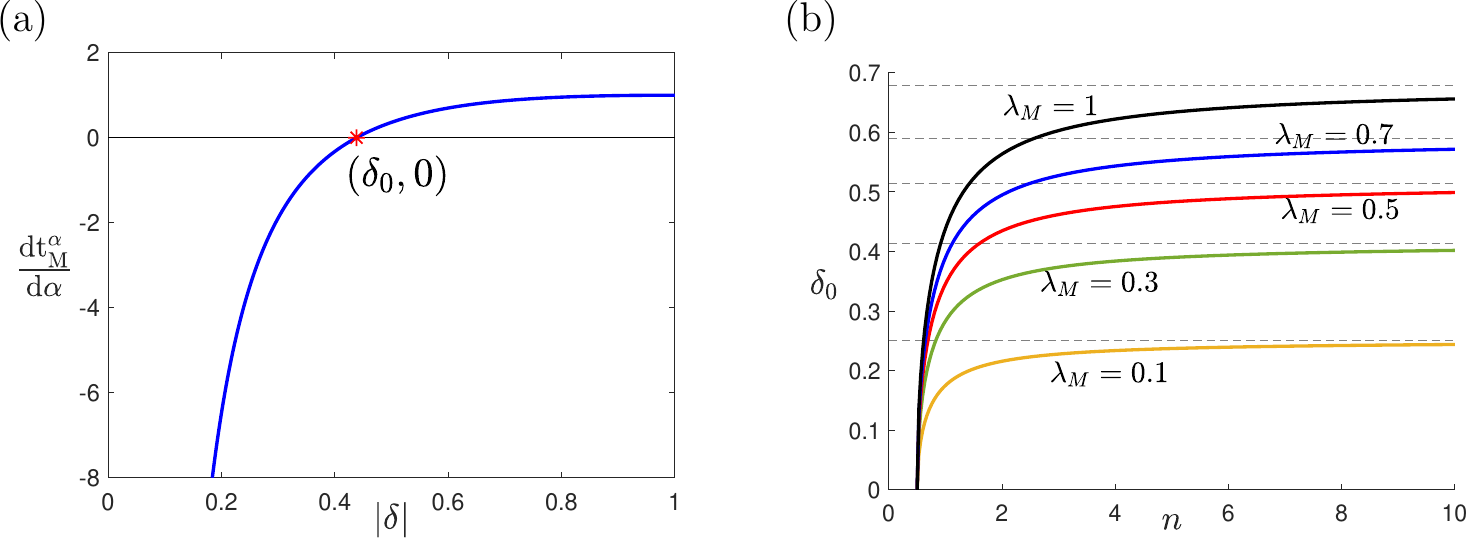}
\caption{ Gent model \eqref{Gent_Stress}: (a) Plot of $\dv{t^{\alpha}_M}{\alpha}$ versus the magnitude of the fractional shear strain for $\lambda_M,n=1$ using \eqref{Gdelta_Gent}. The nature of the plot remains the same for any $\lambda_M(=1 - 1/M) \in (0,1)$ and any $n > 0.5$, as $\delta \to 0^-$, the limit $\dv{t^{\alpha}_M}{\alpha}\to -\infty$  and $\dv{t^{\alpha}_M}{\alpha}$ changes sign from negative to positive at the root $|\hat{\delta}(\alpha)| = \delta_0$.  (b) Plots of $\delta_0$ versus the loading exponent for different values of $\lambda_M$. 
The dotted lines represent the asymptotic value of $\delta_0$ as $n\to \infty$ and $\delta_0 \to 0$ as $n \to 0.5$.}
\label{fig:gent_minimisation}
\end{figure} 
Thus, from \eqref{slopeorder}, $p_0=1, c_0 = {3 c_s}/{J_m} $   for \eqref{gentdcdg}, and from \eqref{n_crit} we have the value of critical loading exponent as $n_c =0.5$.  Thus, for a Gent solid, even though the initial loading rate is infinite for $0.5<n<1$, the limit of acceleration magnification does not become infinite immediately. It does however for $n<0.5$.  
  For loading exponent values $n \ge 0.5$, we have to perform the minimization \eqref{tm} to find $\alpha_{\min}^M$ needed in \eqref{specialisedXmGent}. 
Writing the different functions in terms of fractional strain $\delta$ and then plugging them into \eqref{dtmdalpha} we end up with (see \ref{app:gent})
\begin{align}
\dv{t^{\alpha}_M}{\alpha} = \tilde{G}(\delta) &= 1 - \cfrac{3(1 - \delta^2)^2\lambda_M}{\delta^2(3+\delta^2)^2}  -\cfrac{{{(1+\delta^2}})^{\frac{3}{2}}}{\delta(3+\delta^2)}\frac{(n-1)\lambda_M}{n\tilde{q}(\delta)} \quad 0 \le |\delta| <1 \label{Gdelta_Gent}
\end{align} 
 \begin{figure}
 \begin{centering}
\includegraphics[width=0.5\textwidth]{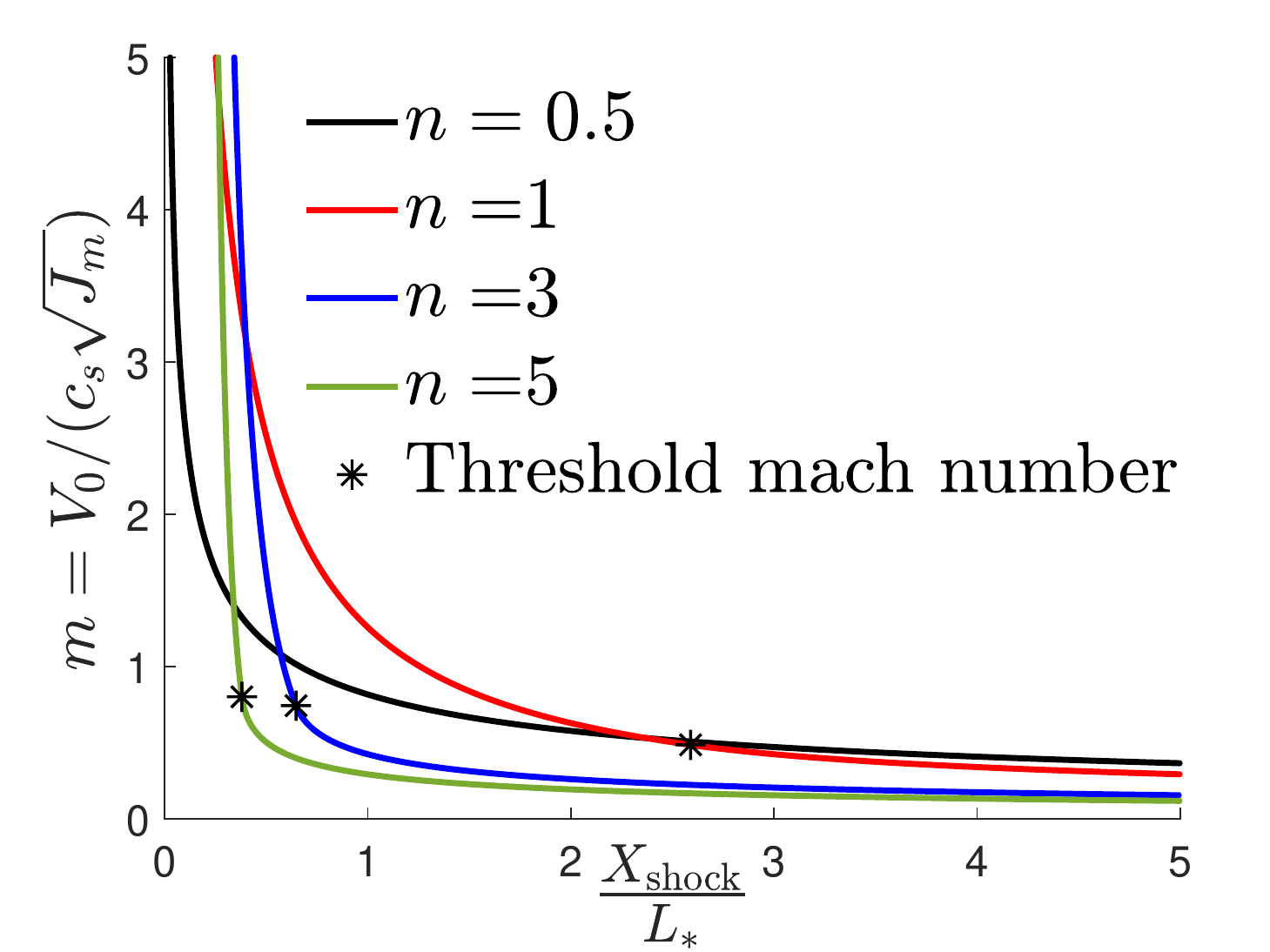}
\caption{Gent model \eqref{Gent_Stress}: Non-dimensional map of the shock location as a function of the loading mach number $m$, for various loading exponents (using $\lambda_M=1$ in \eqref{gent_result}). For $n>0.5(n_c)$, a threshold mach number ($m_\textrm{th}$) separates two different functional forms of $X_\textrm{shock}/L_*$. In different loading mach number regimes different loading exponents might delay shock formation.} \label{fig:Gent_Shock}
\end{centering}
\end{figure}
For $n=0.5 (n_c)$, it can be shown that for $0<|\delta|<1$, the function $\tilde{G}(\delta)$ is positive and hence, 
 $\alpha^M_\textrm{min} = \delta^M_\textrm{min} = 0$. Thus, acceleration magnification and shock formation initiates at the leading edge of the waveform once again for $n=n_c$,  and using \eqref{Xm} and  $\eqref{tm0_ncrit}^2$ we have
\begin{equation}
X_M = \eval{X_M^\alpha}_{\alpha^M_\textrm{min}} =  X^{0}_{M} \eval{}_{n = n_c} =   \frac{2c_s^{3} \ t_0 \lambda_M J_m }{ { 3 V_0^{2}} } =  \frac{2 L_* \lambda_M}{ { 3 m^{2}} } \label{gentXmncrit}
\end{equation}

 For $n>0.5$, the plot of $\dv{t^{\alpha}_M}{\alpha}$ vs $|\delta|$ is qualitatively similar to the plot in Fig. \ref{fig:gent_minimisation}(a), with a zero, $\delta_0 \in (0,1)$. That is, for $|\delta|<\delta_0$, the derivative $\dv{t^{\alpha}_M}{\alpha}$ is negative, for $|\delta|>\delta_0$, it is positive and is zero at $|\delta|= \delta_0$.  Thus if the loading mach number is high enough such that a fractional shear strain magnitude of $\delta_0$ is realized at the loading surface during the loading, then ${\delta_{\textrm{min}}^M}$ will be\footnote{Minus sign since $\delta_0>0$, ${\delta_{\textrm{min}}^M}<0$.}$-\delta_0$. We can evaluate the threshold mach number $m_\textrm{th}$, which is the minimum value of the loading mach number required to realize a fractional strain magnitude of $\delta_0$ during loading,  by setting $m_{\textrm{th}} = \tilde{q}(\delta_0)$ using \eqref{qsolGent} (note that $\delta_0>0$). When $m < m_\textrm{th}$, the loading is not high enough to generate fractional strain magnitudes greater than or equal to $\delta_0$ and since $\dv{t^{\alpha}_M}{\alpha}<0$ for $|\delta|<\delta_0$, the minimizer ${\delta_{\textrm{min}}^M}$ would simply be the largest (in magnitude) fractional strain realized. This would correspond to the strain state at the end of ramping when $|q|$ reaches $m$ in \eqref{qsolGent}. Thus,
  \begin{equation}
 \textrm{for } n>0.5 \quad , \quad {\delta_{\textrm{min}}^M} = - \begin{cases}
 \delta_0  & m \ge m_{\textrm{th}}(=\tilde{q}(\delta_0))  \\ 
\tilde{q}^{-1}(m) & m < m_{\textrm{th}}  \end{cases} \label{deltamingent2}
\end{equation}
Hence, once again, for $n>n_c$, depending on the loading mach number and threshold mach number, earliest realization of a shock or given acceleration magnification can happen at the trailing edge ($\hat{\delta}(t_0) = -\tilde{q}^{-1}(m)$) or the middle of the ramping part of the waveform. Plots of $\delta_0$ as a function of the loading exponent $n$ for different values of $\lambda_M$ are shown in Fig. \ref{fig:gent_minimisation}(b), it can be seen that $\delta_0 \to 0$ as $n \to 0.5$. 
Also, $\delta_0$ is higher for higher acceleration magnification for a given loading exponent $n>0.5$,  and for a given acceleration magnification, $\delta_0$  is higher for higher $n$.

    \begin{figure}[!ht]
\includegraphics[width=\textwidth]{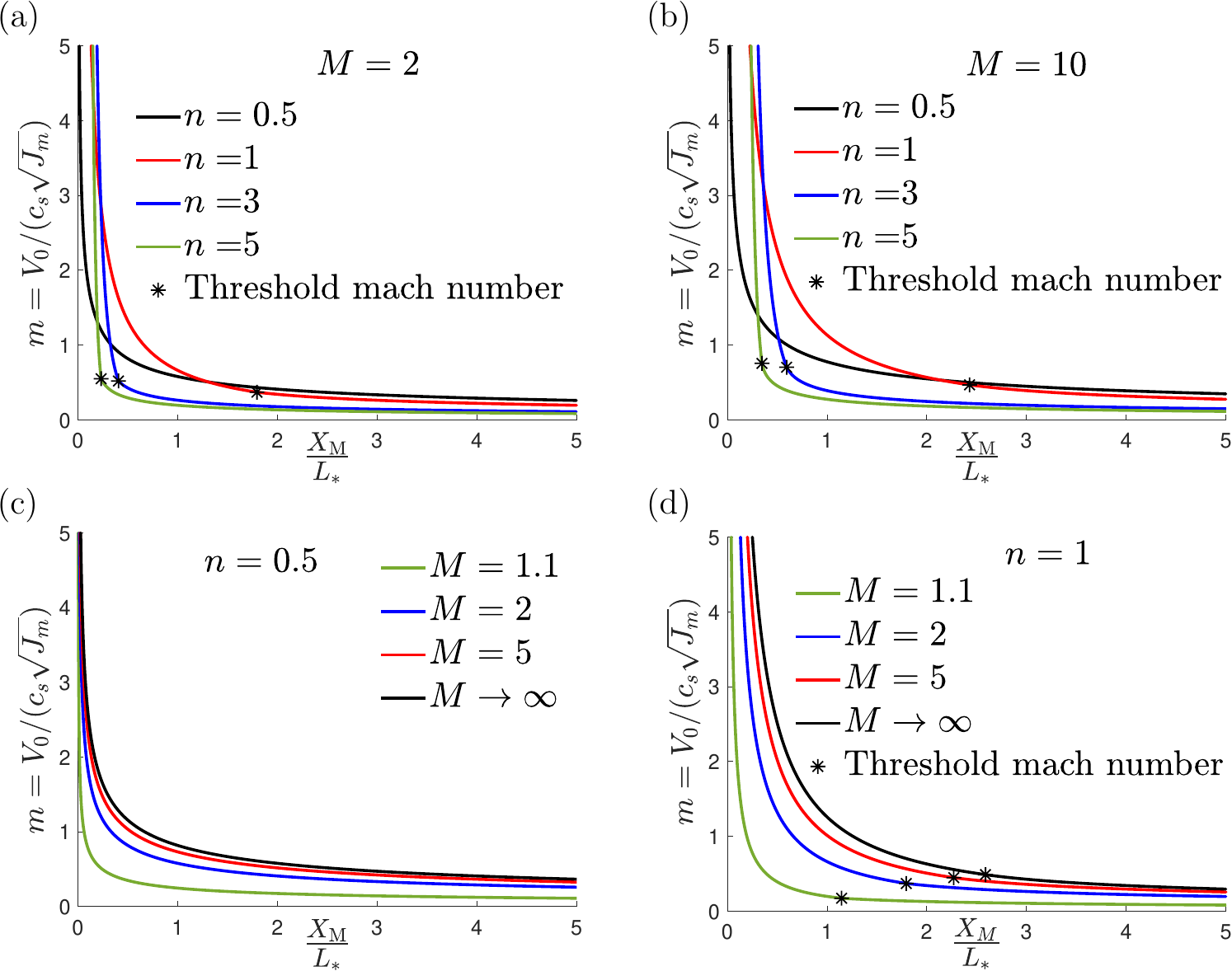}
\caption{Gent model \eqref{Gent_Stress}: Maps of the dimensionless distance of first realization of acceleration magnification $M$, for varying loading mach number $m$, using \eqref{gent_result}. Maps for different loading exponents $n$ with (a) $M=2$, and (b) $M=10$. Maps at  different $M$ for (c) $n=0.5$($n_c$) and (d) $n=1$. In (c),(d), curves for $X_M$ quickly converge to that of the shock  location $(M\to\infty)$ at higher values of $M$. }
\label{fig:gentresult2}
\end{figure}
Plugging \eqref{deltamingent2} into \eqref{specialisedXmGent2}, and using the result \eqref{gentXmncrit}, we obtain the solution for non dimensionalized $X_M$ for Gent solids (see \ref{app:gent}), 
  \begin{equation}\label{gent_result}
\frac{X^M}{L^*} = {\lambda_M}\begin{cases}
\phantom{0}0 & n <0.5,\ m\ge0 \\[3pt] 
\cfrac{2}{3m^2} &   n = 0.5,\ m\ge0\\[11pt] 
\cfrac{K(-\delta_0(\lambda_M,n))}{n\ m^{\frac{1}{n}}\ m_{\textrm{th}}^\frac{n-1}{n}}   & n> 0.5,\ m \ge m_{\textrm{th}}\\[17pt]  
\cfrac{\eval{{K}}_{-\tilde{q}^{-1}(m)}}{n \ m}  & n > 0.5,\ 0 \le m < m_{\textrm{th}} 
 \end{cases} 
\end{equation} 
where $m_{\textrm{th}}=\tilde{q}(\delta_0(\lambda_M,n))$ and $\delta_0(\lambda_M,n)$ is the positive root of $\tilde{G}(\delta)$ in \eqref{Gdelta_Gent}. Setting $\lambda_M =1$ gives the expressions for ${X_{\textrm{shock}}}/{L^*}$ for the incompressible Gent hyperelastic solid. Once again it can be verified that the only discontinuity in the solution space is in moving from $n \to n_c{}^-$ to $n_c$ ($0.5$ for Gent solid). 

 The results for a Gent solid are shown in Figs. \ref{fig:Gent_Shock}-\ref{fig:gentresult2} and can be seen to be exactly similar in nature to the results for the exponential stress model except the fact that $n_c$ is now $0.5$ instead of $1$ and thus shocks are immediately formed for loadings with loading exponent $n<0.5$ and the threshold mach number separating different functional forms of the non-dimensional evolution distances comes in the picture for $n>0.5$. Increasing material nonlinearity (lower $J_m$ instead of higher $k$) once again hastens shock evolution and other observations made for the results of the exponential model carry over. 

\subsection{Ogden model}    
    \begin{figure}
\includegraphics[width=\textwidth]{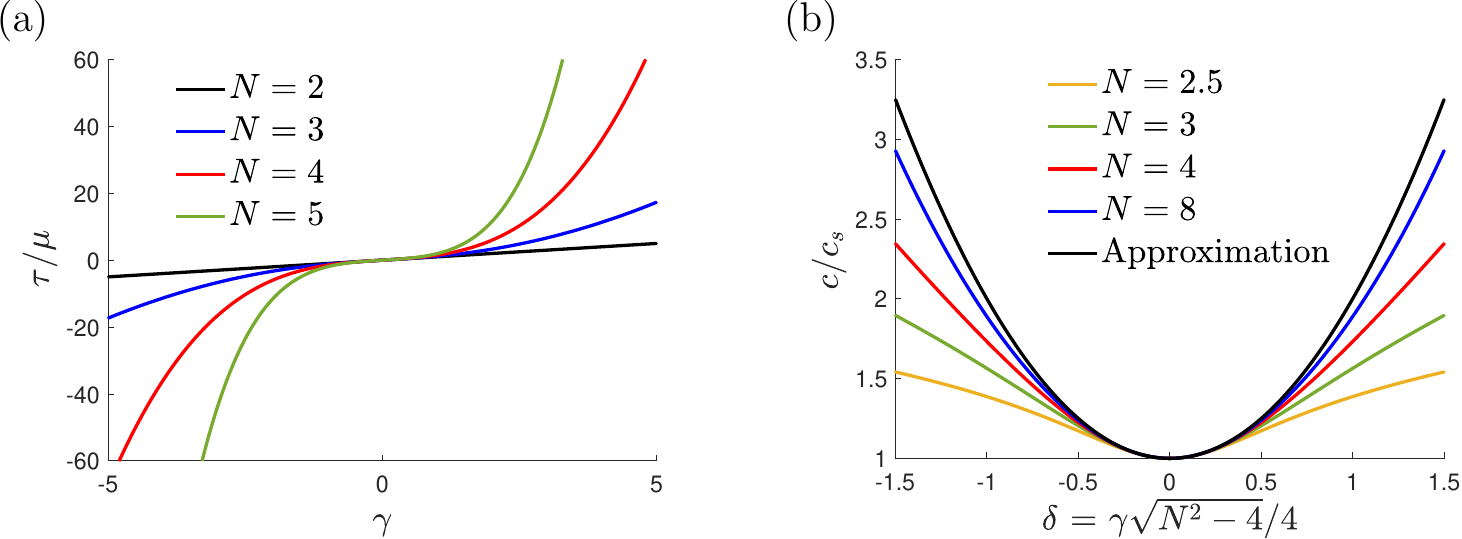}
\caption{Ogden model \eqref{stress_ogden}: (a) Plot of the dimensionless shear stress as a function of the shear strain $\gamma$. The stress response is stiffening for $N>2$ while the value $N=2$ recovers the neo-Hookean limit.  (b) Plot of  the dimensionless shear wavespeed as a function of the modified shear strain $\delta$.  For moderate strains, the approximation of the shear wavespeed profiles as a function of the modified shear strain collapse on a single curve described by \eqref{approx_c}, which approximates the material response well over larger ranges of the modified shear strain for higher $N$.  } \label{fig:ogden_Stress}
\label{fig:Ogdenshearstress}
\end{figure} 
    Another popular hyperelastic energy function commonly used to model incompressible soft solids is the Ogden model which is a multiparameter model expressed in terms of the principal stretches $\lambda_i$, which are the square root of eigenvalues of $\nten B$, and material constants $\mu_p$ and $N_p$ where $p =1,2,3...,P$. For an incompressible material $\lambda_3 = 1/(\lambda_1 \lambda_2)$ 
    and we have the strain energy density function \citep{ogden1972large},
    \begin{equation}
    W\left( \lambda_1,\lambda_2 \right) = \sum_{p=1}^P \frac{2\mu_p}{N_p^2}\left( \lambda_1^{N_p} + \lambda_2^{N_p} + \lambda_1^{-N_p}\lambda_2^{-N_p} -3 \right) \label{ogden_W}
    \end{equation}
    where the linear elastic shear modulus is given by
    \begin{equation}
    \mu = \frac{1}{2} \sum_{p=1}^{N} \mu_p N_{p} 
    \end{equation}
    The Ogden model allows for a large number of parameters that can be chosen to fit different experimental stress-strain curves. Here we use the two parameter Ogden strain energy function for our analysis, i.e $P=1$, such that $N = N_1$ and $\mu = \mu_1 N/2$ are the two independent material parameters that qualify the stress response of the material. 
The nonlinearity of the shear response is captured by the stiffening parameter $N$ and the shear response is given by 
 (using \eqref{ogden_W} in \eqref{Piola_incom})  
\begin{equation}
\tau(\gamma) = \cfrac{2\mu}{N}\cfrac{(\lambda_1^{N}-\lambda_1^{-N})}{\sqrt{4+\gamma^2}}\quad \textrm{where}\quad \lambda_1  = \cfrac{\gamma+\sqrt{\gamma^2+4}}{2} \label{stress_ogden}
\end{equation}    
The sign of $N$ makes no difference to the shear stress response and hence we consider $N>0$ here without loss of generality. Plots of the shear stress as a function of the shear strain are shown in Fig. \ref{fig:ogden_Stress}(a),  for different values of the material parameter $N$. The value of $N = 2$ yields the neo-Hookean shear response. We restrict our attention to materials with $N>2$ for which the response is stiffening. The shear wavespeed for \eqref{stress_ogden} can be found using \eqref{waveeqn}, as
\begin{equation}
c(\gamma) = c_s \sqrt{ \frac{2}{4+\gamma^2}\bigg(\lambda_1^N + \lambda_1^{-N} - \frac{\gamma \tau(\gamma)}{2\mu}\bigg)} 
\label{ogdenwavespeed}
\end{equation}

The large deformation Ogden model is not amenable to find closed form solutions using our analysis as the integration of $c(\gamma)$ is not analytically tractable (to find $Q(\gamma)$). With the goal of demonstrating similarity of results to the other stress models we make the assumption of moderate strains (such that $\gamma^2 >> \gamma^4$). This allows us to use series expansions about $\gamma=0$, allowing for closed form expressions with the caveat of loss of accuracy at larger shear strains. Accordingly, the wavespeed \eqref{ogdenwavespeed} and its derivative can be expanded about $\gamma =0$ as
\begin{equation}
c(\gamma) \approx c_s \bigg(1 + \frac{N^2 - 4}{16}\gamma^2\bigg) \quad , \quad \dv{c}{\gamma} \approx c_s\bigg(\frac{N^2 - 4}{8}\bigg)\gamma  \label{approxwavespeed}
\end{equation}    
Once again we assume $V\ge0,\gamma\le0$ and from \eqref{slopeorder} we have $p_0=1, c_0 = c_s (N^2-4)/8$ for \eqref{approxwavespeed}. Therefore the critical loading exponent $n_c$ from \eqref{n_crit} is once again $0.5$, similar to the Gent model. Thus for loading exponents below $0.5$, shocks are immediately formed for the stiffening Ogden solid (this result does not require the assumption of small/moderate strains). Further, $Q(\gamma)$ can be approximated as
\begin{equation}
Q(\gamma) \approx c_s \gamma \bigg( 1 + \frac{\big(N^2 - 4\big)}{48} \gamma^2 \bigg)
\end{equation}

We define a modified strain $\delta = \gamma \sqrt{N^2-4} /4$, and functions of $\delta$ once again have a tilde 
 accent.  Functions of the shear strain $f(\gamma)$ can be written as $\tilde{f}(\delta) = f(4\delta/ \sqrt{N^2 -4})$ and thus the wavespeed from \eqref{approxwavespeed} becomes 
\begin{equation}
\tilde{c}(\delta) \approx c_s ( 1 + \delta^2 ) \label{approx_c}
\end{equation}  
    \begin{figure}
\includegraphics[width=\textwidth]{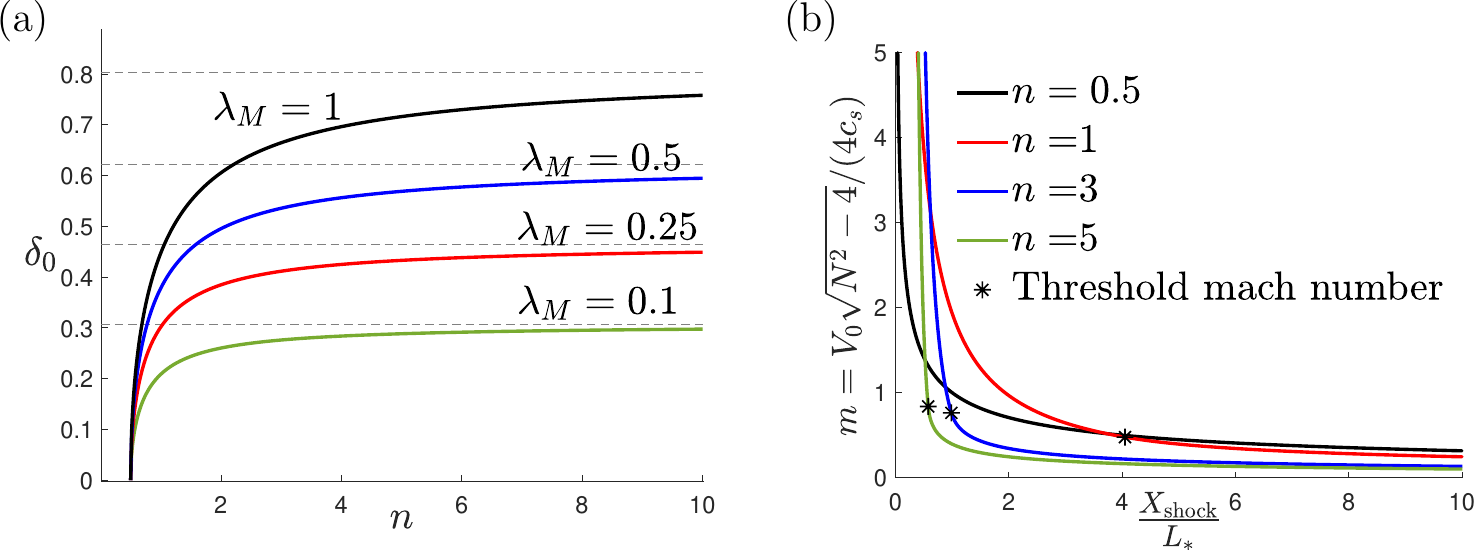}
\caption{ Ogden model \eqref{stress_ogden} for $N>2$: (a) Plots of $\delta_0$, the positive root of \eqref{Gdelta_ogden}, versus the loading exponent for different values of $\lambda_M$. 
The dotted lines represent the asymptotic value of $\delta_0$ as $n\to \infty$ and $\delta_0 \to 0$ as $n\to 0.5 \ (n_c)$. (b) 
Non-dimensionalized map of the shock location as a function of the loading mach number $m$, for various loading exponents (using $\lambda_M=1$ in \eqref{Ogden_result}). For $n>0.5(n_c)$, a threshold mach number ($m_\textrm{th}$) separates two different functional forms of $X_\textrm{shock}/L_*$. In different loading mach number regimes different loading exponents might delay shock formation.
}
\label{fig:Ogdendelta}
\end{figure} 
Plots of the wavespeed as a function of the modified strain for different material stiffening parameters $N>2$ are shown in Fig. \ref{fig:ogden_Stress}(b), at moderate values of $\delta$ all the curves match well with the approximation \eqref{approx_c}. It can be seen that the more the stiffening (larger $N$), the larger the range of modified strain $\delta$ over which the approximation for wavespeed \eqref{approx_c} is good\footnote{At higher $N$ larger range of $\delta$ still corresponds to low/moderate shear strains.}. 

We can use \eqref{Qsolgent} and \eqref{specialisedXmGent} for the Ogden solid as well, except that $\delta$ is now the modified strain and $\tilde{Q}$ is given by
\begin{equation}
\tilde{Q} = V_* \bigg( \delta + \frac{\delta^3}{3} \bigg) \quad \textrm{where} \quad V_* = \frac{4c_s}{\sqrt{N^2-4}} \label{QOgden}
\end{equation}
Equation \eqref{Qsolgent} once again provides a mapping $\hat{F}(\alpha) = \tilde F(\hat{\delta}(\alpha))$. We can divide \eqref{Qsolgent} and \eqref{QOgden} by $V_*$ to write
\begin{equation}
\tilde{q}(\delta) = \frac{Q(\delta)}{V_*} = -\frac{V}{V_*}   \quad \Rightarrow \quad \delta = -\tilde{q}^{-1}\bigg({\frac{V}{V_*}}\bigg) \quad \textrm{where} \quad \tilde{q}(\delta) = \delta + \frac{\delta^3}{3} \label{qsologden}
\end{equation} 
and $|\tilde{q}(\delta)|$ is once again the local mach number, different from the loading mach number $m$. We can specialize \eqref{specialisedXmGent} for the Ogden solid as (see \ref{app:ogden})
\begin{equation}
X_{M} = \cfrac{L_*\lambda_M}{n m^{\frac{1}{n}}} \cfrac{K_\textrm{og}(\delta)}{(-q(\delta))^{\frac{n-1}{n}}} \quad, \quad K_\textrm{og}(\delta) = {\cfrac{-(1+\delta^2)^3}{2\delta}}   \label{specialisedXmOgden2}
\end{equation}
where $m$ and $L_*$ are defined in \eqref{m_andL*}.

Evaluating all the different functions in terms of $\delta$ and substituting them in \eqref{dtmdalpha} we end up with (see \ref{app:ogden})
\begin{align}
\dv{t^{\alpha}_M}{\alpha} = \tilde{G}(\delta) &= 1 + \frac{3\lambda_M}{2} - \frac{\lambda_M}{2\delta^2}  -\cfrac{{3{(1+\delta^2}})^{2}}{2\delta^2(3+\delta^2)}\frac{(n-1)\lambda_M}{n} \label{Gdelta_ogden}
\end{align}  
For $n=0.5$, once again we can verify that $\dv{t_M^\alpha}{\alpha} = \tilde{G}$ is positive for all $\delta$ 
and thus, $\alpha^M_\textrm{min} = \delta^M_\textrm{min} = 0$. 
Hence, acceleration magnification and shock formation initiates at the leading edge of the waveform form for $n=n_c$, as before. For $n = 0.5 (n_c)$, we have using \eqref{Xm} and  $\eqref{tm0_ncrit}^2$
\begin{equation}
X_M = \eval{X_M^\alpha}_{\alpha^M_\textrm{min}} =  X^{0}_{M} \eval{}_{n = n_c} =   \frac{16 c_s^{3} \ t_0  \lambda_M}{(N^2-4) { V_0^{2}} } =  \frac{L_* \lambda_M}{ { m^{2}} } \label{OgdentXmncrit}
\end{equation}

    \begin{figure}[!ht]
\includegraphics[width=\textwidth]{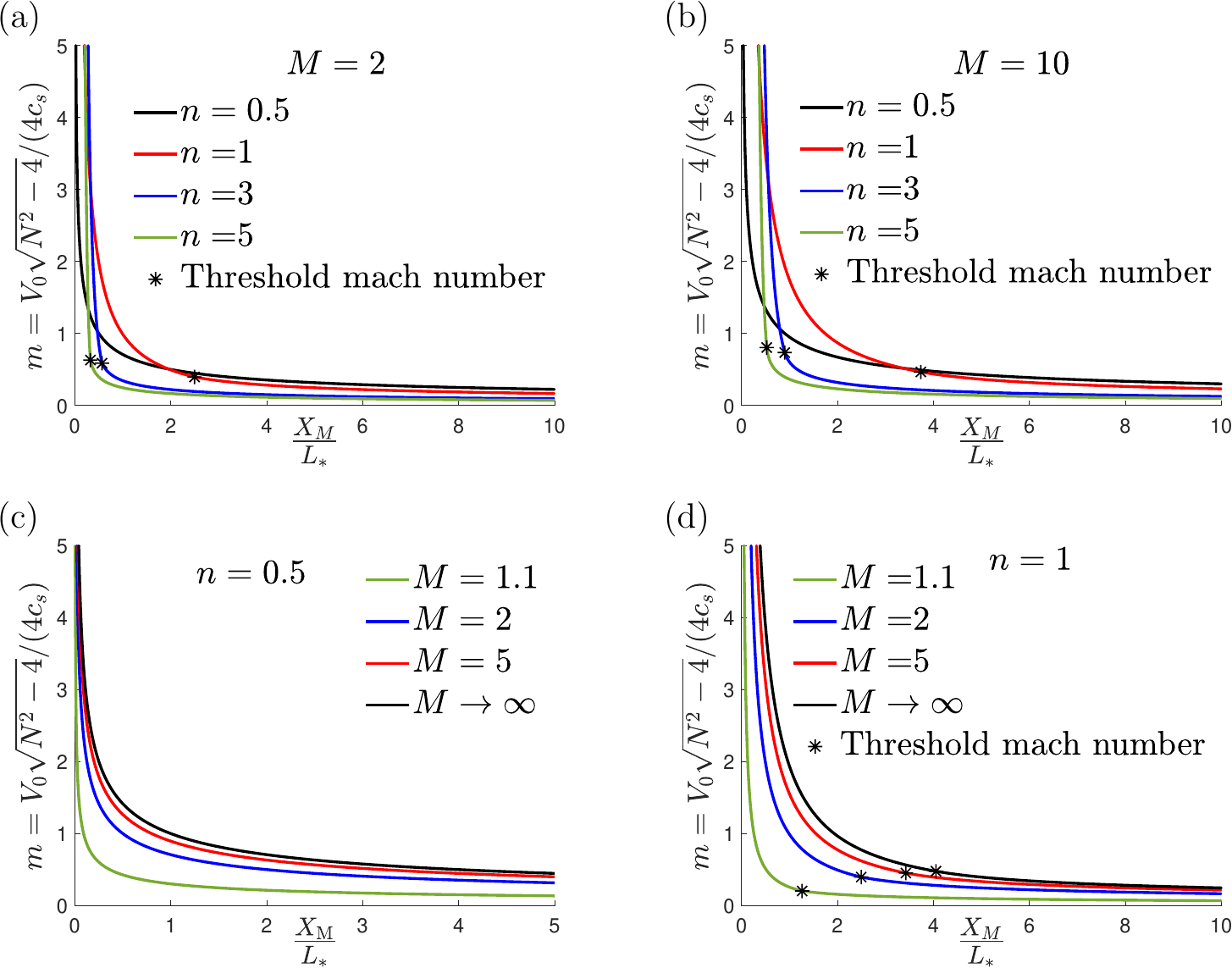}
\caption{Ogden model \eqref{stress_ogden} for $N>2$:  Maps of the dimensionless distance of first realization of acceleration magnification $M$, for varying loading mach number $m$, using \eqref{Ogden_result}. Maps for different loading exponents $n$ with (a) $M=2$, and (b) $M=10$. Maps at  different $M$ for (c) $n=0.5$($n_c$) and (d) $n=1$. In (c),(d), curves for $X_M$ quickly converge to that of the shock  location $(M\to\infty)$ at higher values of $M$. }
\label{fig:Ogden_Xms}
\end{figure}

For $n>0.5$ the plot of $\dv{t^{\alpha}_M}{\alpha}$ vs $|\delta|$ resembles that in Fig. \ref{fig:gent_minimisation}(a), with a zero, $\delta_0 \in (0,1)$. The discussion of $\delta_0$ for the Gent solid holds for the Ogden solid as well  except $\delta$ here refers to the modified strain instead of the fractional shear strain and Fig. \ref{fig:Ogdendelta}(a) shows the plots of $\delta_0$ for an Ogden solid. Similarly 
\eqref{deltamingent2} also applies for the Gent solid where $\tilde{q}$ is now given by \eqref{qsologden}.  Plugging \eqref{deltamingent2} into \eqref{specialisedXmOgden2}, and using the result \eqref{OgdentXmncrit} we obtain the solution for non dimensionalised $X_M$ at moderate strains for stiffening ($N>2$) Ogden solids (see \ref{app:ogden}), 
  \begin{equation} \label{Ogden_result}
\frac{X^M}{L^*} = {\lambda_M}\begin{cases}
\phantom{0}0  & n <0.5,\ m\ge0 \\[3pt]
\cfrac{1}{m^2} &   n = 0.5,\ m\ge0\\[11pt] 
\cfrac{K_{\textrm{og}}(-\delta_0(\lambda_M,n))}{ n\ m^{\frac{1}{n}} \ m_{\textrm{th}}^\frac{n-1}{n}}  & n> 0.5,\ m \ge m_{\textrm{th}}\\[17pt]
  \cfrac{\eval{{K_{\textrm{og}}}}_{-\tilde{q}^{-1}(m)}}{n \ m}  & n > 0.5,\ 0 \le m < m_{\textrm{th}}
 \end{cases}  
\end{equation} 
Here $m_{\textrm{th}}=\tilde{q}(\delta_0(\lambda_M,n))$ and $\delta_0(\lambda_M,n)$ is the positive root of $\tilde{G}(\delta)$ in \eqref{Gdelta_ogden}. Setting $\lambda_M =1$ gives the expressions for ${X_{\textrm{shock}}}/{L^*}$ for the incompressible stiffening Ogden solid. Once again it can be verified that the only discontinuity in the solution space is in moving from $n \to n_c{}^-$ to $n_c$ ($0.5$ for Ogden solid). The results for a stiffening Ogden solid (at moderate strains) are shown in Fig. \ref{fig:Ogdendelta}(b) and Fig. \ref{fig:Ogden_Xms}, and are  similar in nature to the results for the Gent model. A larger $N$ corresponds to higher material nonlinearity similar to a larger $k$ in the exponential stress model and smaller $J_m$ in the Gent model. All the observations made for the results of the Gent model also hold for the stiffening Ogden model when the strains are moderate ($\gamma^2 >> \gamma^4$). The general nature of results might hold for large strains as well though closed form solutions were not attainable here.     
    
\section{Example application to shear shock formation in brain tissue} 
\label{sec:Application}  
    \begin{figure}[htb]
    \begin{center}
\includegraphics[width=0.5\textwidth]{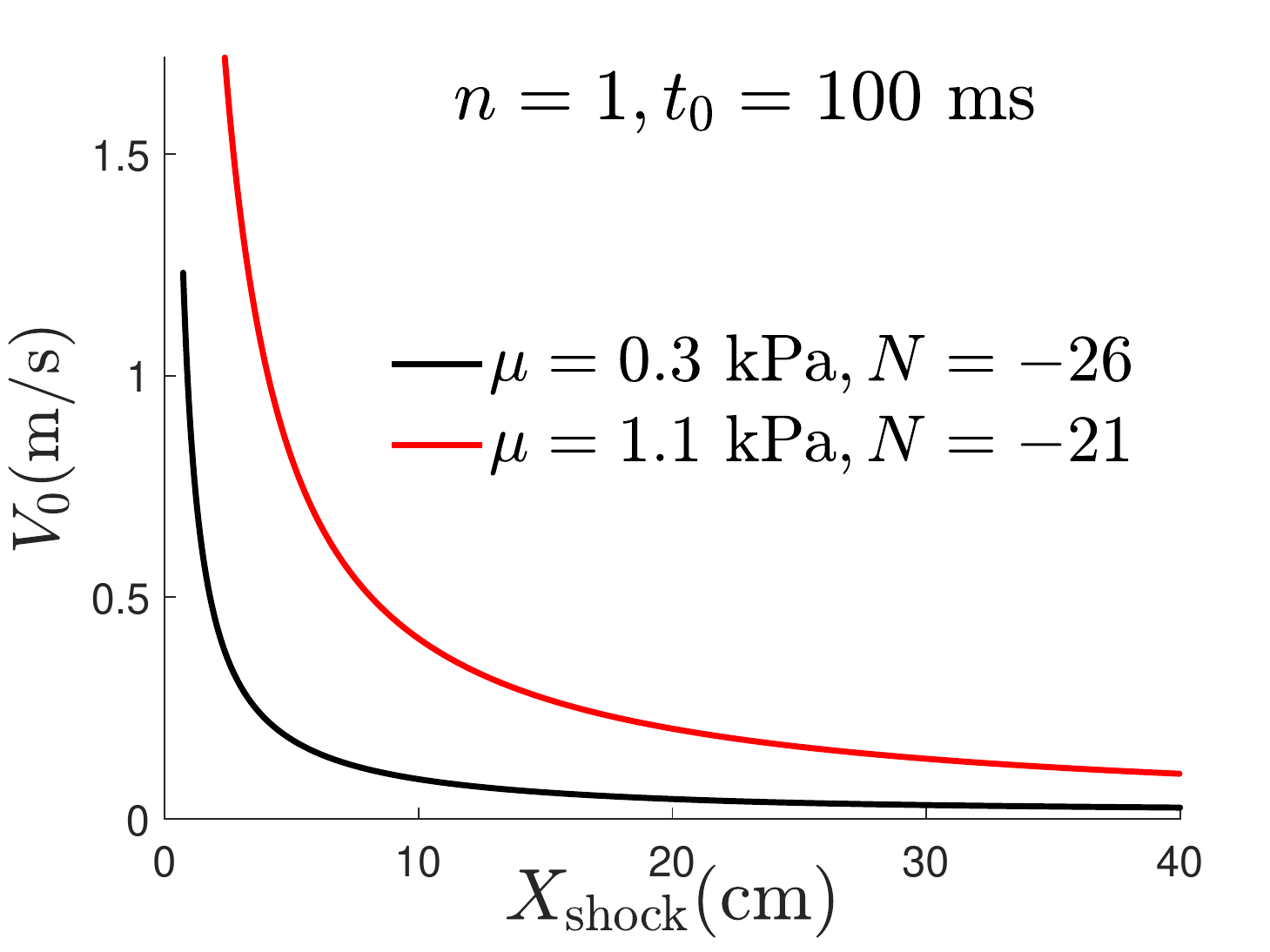}
\caption{Shock formation in the brain: Plot of the shock formation distance for the two parameter Ogden model \eqref{stress_ogden}, as a function of the ramp shear velocity $V_0$, for a linear ramping ($n=1$ in the loading \eqref{loadprofile}) and ramp time of $100$ ms, using \eqref{Ogden_result}. The two curves shown are the maximum and minimum shock formation distances for the range of Ogden parameter values reported in \cite{budday2017mechanical} for simple shear loading of the human brain. For the range of $V_0$ shown, the magnitude of shear strains generated is below $0.5$.  } 
\label{fig:brain_shock}
\end{center}
\end{figure}   

Having quantified shear shock evolution for different hyperelastic models in the previous section, we now apply our results to a physical problem of interest. Recently, \cite{espindola2017shear} demonstrated the evolution of smooth shear waves into shear shocks in a porcine brain, showing an acceleration amplification factor of up to $M\approx8.5$ at the shock front; the magnification is not infinite as in our study, since the brain tissue has some viscosity. Nonetheless, it was seen in the previous section that high magnifications of that order are attained practically close to the shock location. In \cite{espindola2017shear}, it was suggested that shear shock waves could be a previously unappreciated mechanism that could play a significant role in traumatic brain injuries (TBI).  To investigate shear shock formation in the brain using our results, we idealize the brain as an isotropic hyperleastic body and neglect boundary effects and reflections. We also ignore viscosity and any  spatial variation in properties.  The Ogden model is the most commonly used hyperelastic model for constitutive studies of the brain and consequently we will use our results for the two parameter Ogden model.

 The mechanical response of the human brain was carefully and systematically studied across different brain regions for various loading modes in \cite{budday2017mechanical} and it was demonstrated that popular hyperelastic models are unable to accurately capture the stress response across different loading modes with a single set of parameters. However, we are primarily interested in the shear response of the brain and we make use of the parameters reported for the simple shear testing. For the two parameter Ogden shear response \eqref{stress_ogden},  averaged parameter values for $\mu$ in the range of $0.3-1.1\ $kPa and for $N$ in the range of $-26$ to $-21$ were reported for different regions of the brain. We take the density of the brain to be $\rho_0 = 1000\ \textrm{kg/}\textrm{m}^3$ which yields $c_s$ in the range of $0.5-1.1$ \textrm{m/s}. Assuming that a shear impact of the brain can be represented by the loading \eqref{loadprofile}, we consider a linear ramping ($n=1$) from zero velocity to an impact velocity $V_0$, over an impact time of $t_0 =100$ ms. Using these loading and material parameters in \eqref{Ogden_result}, we can make a plot of the shock formation distance as a function of the impact velocity as shown in Fig. \ref{fig:brain_shock}. The two curves shown are for the material parameter values, from the range of reported values, that predict minimum and maximum shock distance. Thus, for a given impact we have a range of distances within which a shock is expected to form for an arbitrary brain sample. As discussed in the previous section, a brain that has a lower $\mu$ and higher nonlinearity parameter ($N$), evolves a shock over a shorter distance for a given impact. Since the Ogden parameter values were reported for tests conducted in the strain magnitude range of $0-0.2$ and since our results for the Ogden solid are more accurate for moderate strains we cap the curves at impact velocity values that would generate a shear strain of\footnote{Maximum value of $V_0 = Q(\gamma = 0.5)$.} $0.5$. Shock formation distance for any impact time can be obtained by remembering that the shock formation distance scales linearly with $t_0$.

 We can readily use Fig. \ref{fig:brain_shock} in a design problem such as that of a helmet to avoid shear shocks in the brain. 
Since the average length of the human brain is $15$ cm, from Fig. \ref{fig:brain_shock},  we find a conservative\footnote{Viscosity would allow for higher impact velocities without shock formation.} maximum velocity of $\sim 6$ cm/s that can be safely transferred to the brain after dissipation by the helmet and the skull, to avoid shear shocks. Also, the order of shock distances predicted by our analysis for the brain tissue is in agreement with the experimental values in \cite{espindola2017shear} where shocks (peak acceleration magnification) occur at distances of the order of $1$ cm for shear velocities of the order $1$ m/s. These results also suggest that experiments conducted on small brain samples in TBI research might not be able to capture shear shocks that could evolve in a real brain subjected to the same loading.  Nevertheless, the non-dimensional maps can be used to rescale the loading waveform so that similar features of the shock evolution can be observed at smaller scales.  Finally, it can be clearly seen that larger brains are more susceptible to shear shocks.

\section{Conclusions} 
\label{sec:Conclusions}
The nonlinear evolution of a loading shear wave into a shear shock  was studied for incompressible strain stiffening hyperelastic solids using the method of characteristics in the framework of large deformation elastodynamics. A general family of loading waveforms was considered and closed form solutions for the distance of first shock formation or realization of a given acceleration magnification were obtained for three different stress models. The results were encapsulated in non-dimensional maps, formulated in terms of a  loading mach number and  the loading exponent. The maps can guide design selection of loading, and material parameters and dimensions, to avoid shock formation. The singularity of the loading required to immediately form shocks was quantified through a critical loading exponent. For weaker loadings a critical mach number, dependent on the loading waveform shape exists above and below which the functional forms of the evolution distances are different. Larger loadings over shorter times and/or higher material nonlinearity lead to shock evolution over shorter distances. The dependence of the shock evolution on the load profile shape was shown to be non-trivial and different shapes were seen to be better at delaying shock formation depending on the loading mach regime. 
We hope that our non-dimensional maps can guide design of protective structures and also that this work will serve as a stepping stone for future studies of nonlinear wave evolution in solids in the large deformation elastodynamics framework. Future directions include extension to coupled longitudinal and shear waves in compressible solids, inclusion of viscosity, and studying the evolution post shock initiation.

          
\appendix


\section{Shock condition equivalence}
\label{app:shockcond}
\noindent We assume the shear strain does not change sign. From Sec. \ref{sec:shock_formation}, we have the shock condition $\hat{c}'>0$. Since $c(\gamma)$ is an even function, we have $c(\gamma) = c(|\gamma|)$. Thus we have $\hat{c}' = \pdv{c}{|\gamma|}\dv{|\hat{\gamma}|}{\alpha}$ where the sign of $\dv{|\hat{\gamma}|}{\alpha} = {\dv{|\gamma(0,t)|}{t}}$ is decided by the loading program (positive for loading and negative for unloading) and $\pdv{c}{|\gamma|}$ has the same sign as $\tau''(|\gamma|)$. Thus a strain stiffening material ($\tau''(|\gamma|)>0$) would develop shocks under shear loading and a strain softening material ($\tau''(|\gamma|)<0$) would develop shear shocks in unloading. This is consistent with the discussion in Sec. \ref{subsec:shock_cond}.

\section{Solution for loading \eqref{loadprofile}}    

\subsection{Derivation of $X_M$, ${t^{0}_{M}}$}
\label{app:firstmeet}
\noindent We have from \eqref{tMzero}, 
\begin{equation}
t^{0}_{M}  = - \ c_s^2 \lim\limits_{\substack{\alpha \to 0^+ \\\gamma \to 0^- }}{\bigg( V'(\alpha) \dv{c}{\gamma}\bigg)^{-1}}  \label{appendix_tshockzero}
\end{equation}
For the loading \eqref{loadprofile}, using \eqref{Qsol}
\begin{equation}
Q(\gamma) = -V_0\bigg(\frac{\alpha}{t_0}\bigg)^n \quad,\quad \alpha \le t_0
\end{equation}
Thus 
\begin{equation}
V'(\alpha) = \frac{n V_0}{t_0} \bigg(\frac{\alpha}{t_0}\bigg)^{n-1} = \frac{n V_0}{t_0} \bigg(\frac{-Q(\gamma)}{V_0}\bigg)^{\frac{n-1}{n}} \quad , \quad \alpha < t_0
\label{acctostrain}
\end{equation}
Substituting \eqref{acctostrain} in \eqref{Xm2},  we obtain
\begin{align}
X_{M} &= \lambda_M \eval{\frac{\psi c}{\frac{nV_0}{t_0}\big(\frac{-Q(\gamma)}{V_0}\big)^{\frac{n-1}{n}}}}_{\gamma_{\textrm{min}}^M} = \frac{ t_0 \lambda_M}{n V_0^\frac{1}{n}} \eval{\frac{\psi c}{(-Q(\gamma))^{\frac{n-1}{n}}}}_{\gamma_{\textrm{min}}^M} \quad,\quad \gamma_{\textrm{min}}^M = \hat{\gamma}(\alpha_{\min}^M) 
 \label{app:specialised_Xm}
\end{align}
Substituting \eqref{acctostrain} in \eqref{appendix_tshockzero} and using  $Q(\gamma) \to c_s \gamma$ as $\gamma \to 0$, we can write (we assume $V\ge0, \gamma\le0$),
\begin{equation}
t^{0}_{M} =  \frac{c_s^{1+\frac{1}{n}} \ t_0 \lambda_M}{ {n V_0^{1/n}} } \lim_{\gamma \to 0^-} -\bigg(\dv{c}{\gamma}\bigg)^{-1} (-\gamma)^{\frac{1-n}{n}} 
\end{equation}

\subsection{Derivation of ${\dv{t^{\alpha}_M}{\alpha}}$}
\label{app:derivativetm}
   \noindent  We have from \eqref{tm1}
\begin{equation}
t_{M}^\alpha = \alpha + \lambda_M \frac{\hat{\psi}}{V'(\alpha)}  
\end{equation}
Thus
\begin{equation}
\dv{t^{\alpha}_M}{\alpha} = 1+ \frac{\hat{\psi}'}{V'(\alpha)}\lambda_M - \frac{\hat{\psi}V''(\alpha)\lambda_M}{(V'(\alpha))^2}
\end{equation}
For $\alpha \in [0,t_0]$ and the loading \eqref{loadprofile}, using  \eqref{Qsol} we can write
\begin{equation}
\frac{V''(\alpha)}{(V'(\alpha))^2} = \frac{n-1}{n V(\alpha)} = -\frac{n-1}{n \hat{Q}} \quad,\quad \alpha \in [0,t_0]  \label{app:dummy10}
\end{equation}
We also have using \eqref{Vtogammarate},
\begin{equation}
\frac{\hat{\psi}'}{V'(\alpha)} = \frac{\dv{\psi}{\gamma}\hat{\gamma}'}{V'(\alpha)} = -{\dv{\psi}{\gamma}}{c^{-1}}
\end{equation}
Thus the derivative $\dv{t^{\alpha}_M}{\alpha}$ for $\alpha \in [0,t_0]$ is ,
\begin{equation}
\Rightarrow \dv{t^{\alpha}_M}{\alpha} = \hat{G}(\alpha) = 1+ \hat{R}\lambda_M + \frac{\hat{\psi}(n-1)\lambda_M}{\hat{Q}n} \quad , \quad {R}(\gamma) = -{\dv{\psi}{\gamma}}{c^{-1}} 
\end{equation}
Using $\psi(\gamma) = -{c^2}\big({\dv{c}{\gamma}}\big)^{-1}$, we have ${R}(\gamma) = -{\dv{\psi}{\gamma}}{c}^{-1} = 2 - c{\dv[2]{c}{\gamma}}{(\dv{c}{\gamma})^{-2}}$ and thus
\begin{equation}
G(\gamma) = 1+ 2\lambda_M  - \frac{c \lambda_M }{(\dv{c}{\gamma})^2}\dv[2]{c}{\gamma}   - \frac{(n-1)\lambda_M}{n } \frac{c^2}{Q\dv{c}{\gamma}} \label{app:dtmdalphatwo}
\end{equation}

\section{Application to material models} 

\subsection{Exponential model}
\label{app:exp}
\smallskip
\noindent In this section $\gamma <0$,
\begin{equation}
c(\gamma) = c_s e^{-\frac{k}{2}{\gamma}} \quad ,\quad \dv{c}{\gamma} = - \frac{c_s k}{2} e^{-k\frac{{\gamma}}{2}} \quad,\quad \psi(\gamma)  = V_* e^{-k\frac{{\gamma}}{2}} \quad,\quad {R}(\gamma) =  -{\dv{\psi}{\gamma}}{c}^{-1} = 1 \label{app_exp1}
\end{equation}
From \eqref{gammaexp} we have $e^{-\frac{k}{2}\gamma} = 1 + {V}/{V_*}$, which along with \eqref{app_exp1} yields \eqref{exp_gamma_funcs}.
\smallskip

\noindent \textbf{Evaluation of $\boldsymbol{\alpha_{\textrm{min}}^M}$ for $\boldsymbol{n>1}$:} Differentiating \eqref{tderiv_exp} we have
\begin{equation}
\dv[2]{t^{\alpha}_M}{\alpha} = \frac{(n-1)\lambda_M}{nV^2} V'(\alpha) \label{app_exp2}
\end{equation}
Substituting the loading \eqref{loadprofile} for $\alpha \le t_0$ in \eqref{app_exp2} we obtain
\begin{equation}
\dv[2]{t^{\alpha}_M}{\alpha} = \frac{(n-1)\lambda_M V_*}{\alpha V} 
\end{equation}
Thus $\dv[2]{t^{\alpha}_M}{\alpha} > 0$ for $n > 1$ and $ \alpha \le t_0$, and thus we can find $\alpha_{\textrm{min}}^M$ using \eqref{tderiv_exp} and \eqref{loadprofile} as
\begin{equation}
\eval{\dv{t^{\alpha}_M}{\alpha}}_{\alpha_{\textrm{min}}^M} = 0 \quad \Rightarrow \quad \cfrac{V(\alpha_{\textrm{min}}^M)}{V_*} = m \bigg(\frac{\alpha^M_\textrm{min}}{t_0}\bigg)^n = \frac{(n-1)\lambda_M}{n+\lambda_M} = m_{\textrm{th}} \quad , \quad  \alpha_{\textrm{min}}^M \in [0,t_0] \label{expminalpha1}
\end{equation} 
Equation \eqref{expminalpha1} has a solution $\alpha_{\textrm{min}}^M \in [0,t_0]$ for $n>1$ only if the loading mach number $m$ is greater than or equal to  $m_{\textrm{th}}$. For $m < m_{\textrm{th}}$, $\alpha_{\textrm{min}}^M = t_0$ since $\dv{t^{\alpha}_M}{\alpha} <0$ for $\alpha$ $\in$ $[0,t_0]$ (Using $V/V_* < m < m_\textrm{th}$ in \eqref{tderiv_exp}). Using \eqref{expminalpha1}, for $m \ge m_{\textrm{th}}$,
\begin{equation}
 \alpha^M_\textrm{min} = t_0 \bigg( \frac{m_\textrm{th}}{m} \bigg)^\frac{1}{n} 
\end{equation}

\noindent \textbf{Evaluation of $\boldsymbol{X_M/L_*}$ for $\boldsymbol{n>1}$:} 
Substituting \eqref{exp_gamma_funcs} and the loading \eqref{loadprofile} for $t<t_0$ in \eqref{Xm2} we obtain
\begin{align}
X_{M} = c_s t_0^n \lambda_M \ \cfrac{(V(\alpha_{\min}^M) + V_*) \big(1+\frac{V(\alpha_\textrm{min})}{V_*} \big) }{n V_0 (\alpha_{\min}^M)^{n-1} } = c_s t_0^n \lambda_M \ \cfrac{(\frac{V(\alpha_{\min}^M)}{V_0} + \frac{1}{m}) \big(1+\frac{V(\alpha^M_\textrm{min})}{V_*} \big) }{n  V_0(\alpha_{\min}^M)^{n-1} }\label{app:Xmexp}
\end{align}
\noindent From \eqref{expalphamins} and \eqref{loadprofile}, for $n>1$ and $m<m_\textrm{th}$, we have $\alpha_{\min}^M = t_0, V(\alpha_{\min}^M) = V_0$, and thus \eqref{app:Xmexp} becomes
\begin{equation}
X_{M} = \frac{c_s t_0 \lambda_M}{n} \ {\bigg(1 + \frac{1}{m}\bigg) ( 1+m ) }  \Rightarrow \cfrac{X_{M}}{L_*} = \cfrac{\lambda_M}{n}\bigg(2+m+ \frac{1}{m}\bigg) \quad \textrm{for } n>1,{ m<m_\textrm{th}}
\end{equation}
From \eqref{expalphamins} and \eqref{loadprofile}, for $n>1$ and $m\ge m_\textrm{th}$ we have $\alpha_{\min}^M = t_0 \bigg(\cfrac{m_\textrm{th}}{m}\bigg)^{\frac{1}{n}}$, $V(\alpha_{\min}^M)/V_0 = m_\textrm{th}/m$, and thus \eqref{app:Xmexp} becomes
\begin{align}
&X_{M} = L_*\cfrac{ \big(1+m_\textrm{th} \big)^2 \bigg( \cfrac{m_\textrm{th}}{m}\bigg)^{\frac{1}{n}} }{n \ {m_\textrm{th}} } \Rightarrow \cfrac{X_{M}}{L_*} = \cfrac{n(1+\lambda_M)^2}{(n-1)(n+\lambda_M)} \bigg(\cfrac{m_{\textrm{th}}}{m}\bigg)^{\frac{1}{n}} \quad \textrm{for } n>1, m\ge m_\textrm{th}
\end{align}
where we have used $m_{\textrm{th}} = \frac{(n-1)\lambda_M}{n+\lambda_M}$. 
\smallskip

\noindent \textbf{Solution continuity: } We demonstrate the continuity of the solution \eqref{Xmsexp} at $n \to 1^+$ and $m \to m_{\textrm{th}}^-$. We have $m_{\textrm{th}} = \frac{(n-1)\lambda_M}{n+\lambda_M}$ and thus for $n \to 1$, $m_{\textrm{th}} \to 0$. This means that only the $m\ge m_{\textrm{th}}$ case for $n>1$ is relevant for $n \to 1^+$ in \eqref{Xmsexp}. Setting $n \to 1^+$ for that case recovers the $n=1$ case,
\begin{equation}
\frac{X_M}{L_*} = \lim_{n\to1^+} \cfrac{n(1+\lambda_M)^2}{(n-1)(n+\lambda_M)} \bigg(\cfrac{(n-1)\lambda_M}{(n+\lambda_M) m}\bigg)^{\frac{1}{n}} = \frac{\lambda_M}{m}
\end{equation} 
Setting $m = m_{\textrm{th}}$ in the $n\ge1, m\ge m_\textrm{th}$ case in \eqref{Xmsexp} yields
\begin{equation}
\frac{X_M}{L_*} = \cfrac{n(1+\lambda_M)^2}{(n-1)(n+\lambda_M)}  \label{app_exp10}
\end{equation}
Setting $m \to m_{\textrm{th}}^- $ in the $n\ge1, m< m_\textrm{th}$ case in \eqref{Xmsexp} recovers \eqref{app_exp10},
\begin{equation}
\frac{X_M}{L_*} =  \cfrac{\lambda_M}{n}\bigg(\frac{(m_{\textrm{th}} +1 )^2}{m_{\textrm{th}}}\bigg) = \cfrac{n(1+\lambda_M)^2}{(n-1)(n+\lambda_M)} 
\end{equation} 
\noindent \textbf{Shortest distance of realization of acceleration magnification: } 
Let us denote the  shortest distance from the loading surface at which a given magnification $M$ is realized by $X^s_M$.
For $n < n_c$, $X_M$ and $X^s_M$ are both zero. For $n>n_c, m<m_\textrm{th}$, the derivative $\dv{t^\alpha_M}{\alpha}$ is always negative in the ramping interval (Using $V/V_* < m < m_\textrm{th}$ in \eqref{tderiv_exp}) and thus using \eqref{dXalphaM_todtalphaM}, $\dv{X^\alpha_M}{\alpha}$ is also negative during ramping. Thus the characteristic $\alpha = t_0$ will minimize both $t^\alpha_M$ and $X^\alpha_M$ and hence once again $X_M$ and $X^s_M$ are equal. For $n=n_c$, the derivative $\dv{t^\alpha_M}{\alpha}$ is always positive and greater than one (from  \eqref{tderiv_exp}) and hence, using \eqref{dXalphaM_todtalphaM} $\dv{X^\alpha_M}{\alpha}$ is also always positive. Thus the characteristic $\alpha=0$ will minimize both $t^\alpha_M$ and $X^\alpha_M$, and thus $X_M$ and $X^s_M$ are equal once more. For the case $n > n_c, m>m_\textrm{th}$, the distances  $X_M$ and $X^s_M$ will not be the same. The same observations made here can be verified for the Gent and Ogden models as well where $n_c$ will be $0.5$ for them.

\subsection{Gent model}
        \label{app:gent}
        \smallskip
\noindent Evaluating the different expressions for the Gent solid,
\allowdisplaybreaks
\begin{align}
\begin{split}
    c(\gamma) ={}& \sqrt{\frac{1}{\rho_0}\pdv{\tau}{\gamma}} =V_* \cfrac{\sqrt{J_m + \gamma^2}}{J_m - \gamma^2} \quad, \quad \tilde{c}(\delta) = c_s \cfrac{\sqrt{1 + \delta^2}}{1 - \delta^2} \label{app:c_gent}
\end{split}\\
\begin{split}\label{eq:1}
        Q(\gamma)  ={}& V_* \Bigg(\sqrt{2}\tanh^{-1}\Big({\frac{\sqrt{2}\gamma}{\sqrt{J_m+\gamma^2}}}\Big) - \log\Big({\frac{\gamma+\sqrt{J_m + \gamma^2}}{\sqrt{J_m}}    }\Big)\Bigg) \\
        ={}& V_* \log\Bigg({\frac{ \bigg(\frac{\sqrt{J_m + \gamma^2} + \gamma\sqrt{2}}{\sqrt{J_m + \gamma^2} - \gamma\sqrt{2}}\bigg)^\frac{1}{\sqrt{2}}}{\frac{\gamma+\sqrt{J_m + \gamma^2}}{\sqrt{J_m}}}    }\Bigg) \qquad \rm{As \ \gamma}\rightarrow{0}, \quad Q(\gamma) \rightarrow{c_s \gamma }
\end{split}\\
\begin{split}\label{app:Qdeltagent}
           \tilde Q(\delta) ={}&  V_* \log\Bigg({\cfrac{ \bigg(\frac{\sqrt{1 + \delta^2} + \delta\sqrt{2}}{\sqrt{1 + \delta^2} - \delta\sqrt{2}}\bigg)^\frac{1}{\sqrt{2}}}{{\delta+\sqrt{1 + \delta^2}}}}\Bigg)  \qquad \rm{As \ \delta}\rightarrow{0}, \quad \tilde{Q}(\delta) \rightarrow{c_s \sqrt{J_m}\delta }  
\end{split}\\
           \tilde q(\delta) ={}& \frac{Q(\delta)}{V^*}=  \log\Bigg({\cfrac{ \bigg(\frac{\sqrt{1 + \delta^2} + \delta\sqrt{2}}{\sqrt{1 + \delta^2} - \delta\sqrt{2}}\bigg)^\frac{1}{\sqrt{2}}}{{\delta+\sqrt{1 + \delta^2}}}}\Bigg)\label{eq:3}\\
\dv{c(\gamma)}{\gamma} ={}& f_1(\gamma) = \cfrac{c_s\sqrt{J_m}\gamma(3J_m+\gamma^2)}{(J_m - \gamma^2)^2\sqrt{J_m+\gamma^2}} \quad , \quad \tilde{f_1}(\delta) = \cfrac{c_s \delta(3+\delta^2)}{\sqrt{J_m}(1 - \delta^2)^2\sqrt{1+\delta^2}}
\\
\psi(\gamma) ={}& -\frac{c^2(\gamma)}{f_1(\gamma)} = -\cfrac{{V_*{(J_m+\gamma^2}})^{\frac{3}{2}}}{\gamma(3J_m+\gamma^2)} \quad , \quad    \tilde{\psi}(\delta) = -\cfrac{{V_*{(1+\delta^2}})^{\frac{3}{2}}}{\delta(3+\delta^2)} \label{app:pdeltaGent}  \\
\dv{\psi(\gamma)}{\gamma} ={}& -\cfrac{3c_s J_m^{\frac{3}{2}}(\gamma^2-J_m)\sqrt{J_m+\gamma^2}}{\gamma^2(3J_m+\gamma^2)^2}\\
    R(\gamma) ={}&  -{\dv{\psi}{\gamma}}{c}^{-1} = -\cfrac{3J_m (J_m - \gamma^2)^2}{\gamma^2(3J_m+\gamma^2)^2} \quad , \quad 
        \tilde{R}(\delta) = \cfrac{-3(1 - \delta^2)^2}{\delta^2(3+\delta^2)^2} 
\end{align}
Substituting the above expressions in \eqref{dtmdalpha} and using \eqref{qsolGent} we obtain ($\tilde{G}(\delta) = \tilde{G}(\hat{\delta}(\alpha)) = \hat{G}(\alpha)$)
    \begin{align}
\dv{t^{\alpha}_M}{\alpha} = \tilde{G}(\delta) &= 1 - \cfrac{3(1 - \delta^2)^2\lambda_M}{\delta^2(3+\delta^2)^2}  -\cfrac{{{(1+\delta^2}})^{\frac{3}{2}}}{\delta(3+\delta^2)}\frac{(n-1)V_*\lambda_M}{n\tilde{Q}(\delta)}   \label{app:Gdelta_Gent}\\
 &=  1 - \cfrac{3(1 - \delta^2)^2\lambda_M}{\delta^2(3+\delta^2)^2}  -\cfrac{{{(1+\delta^2}})^{\frac{3}{2}}}{\delta(3+\delta^2)}\frac{(n-1)\lambda_M}{n\tilde{q}(\delta)} 
\end{align} 
        
\noindent Using \eqref{app:c_gent} and \eqref{app:pdeltaGent} in \eqref{specialisedXmGent} we have 
\begin{align}
X_{M} &= \frac{\lambda_M t_0}{n V_0^\frac{1}{n}} \eval{\frac{\tilde{\psi}\tilde{c}}{(-\tilde{Q}(\delta))^{\frac{n-1}{n}}}}_{\delta_{\rm{min}}^M} = \frac{c_s t_0 V_*  \lambda_M}{n V_0^{\frac{1}{n}}} \eval{ \cfrac{-(1+\delta^2)^2}{\delta(3+\delta^2)(1-\delta^2)(-Q(\delta))^{\frac{n-1}{n}}} }_{\delta_{\rm{min}}^M}  \\
&= \frac{c_s t_0 V_*^{\frac{1}{n}}  \lambda_M}{n V_0^{\frac{1}{n}}} \eval{ \cfrac{K(\delta)}{(-q(\delta))^{\frac{n-1}{n}}} }_{\delta_{\rm{min}}^M}  = \frac{L_*\lambda_M}{n m^{\frac{1}{n}}}\eval{ \cfrac{K(\delta)}{(-q(\delta))^{\frac{n-1}{n}}} }_{\delta_{\rm{min}}^M} \label{app:gentresult1}
\end{align}  

Using \eqref{deltamingent2} in \eqref{app:gentresult1} we obtain
  \begin{equation}
\frac{X^M}{L^*} = \cfrac{\lambda_M}{nm^{\frac{1}{n}}}\begin{cases}
\cfrac{K(-\delta_0(\lambda_M,n))}{ m_{\rm{th}}^\frac{n-1}{n}}  & m \ge m_{\rm{th}}, n > 0.5 \\ 
\cfrac{\eval{{K}}_{-\tilde{q}^{-1}(m)}}{{m^\frac{n-1}{n}}}  & m < m_{\rm{th}}, n > 0.5 \\
 \end{cases} \label{app:gentresult2}
\end{equation}         

\subsection{Ogden model}
\label{app:ogden}
Under the assumptions of moderate strains, for the Ogden model we have
\begin{align}
\begin{split}
    c(\gamma) \approx{}& c_s \bigg(1 + \frac{N^2 - 4}{16}\gamma^2\bigg)  \quad, \quad \tilde{c} \approx c_s ( 1 + \delta^2 ) \quad , \quad \delta = \frac{\sqrt{N^2 - 4}}{4}\gamma\label{app:c_ogden}
\end{split}\\
\begin{split}
        Q(\gamma)  \approx{}& c_s\gamma(1 + \frac{N^2 - 4}{48}\gamma^3) \quad , \quad Q(\delta) = V_* \delta \bigg(1+\frac{\delta^2}{3}\bigg) \quad , \quad V_* = \frac{4c_s}{\sqrt{N^2-4}} 
\end{split}\\
           \tilde q(\delta) ={}& \frac{Q(\delta)}{V^*}\approx \delta +\frac{\delta^3}{3} \\         
\dv{c(\gamma)}{\gamma} ={}& f_1(\gamma) \approx c_s \frac{N^2 - 4}{8}\gamma \quad , \quad \tilde{f_1}(\delta) =  c_s \frac{\sqrt{N^2 - 4}}{2}\delta \quad , \quad \dv[2]{c(\gamma)}{\gamma} \approx{} c_s \frac{N^2 - 4}{8} 
\\
\psi(\gamma) ={}& -\frac{c^2(\gamma)}{f_1(\gamma)} \approx -\frac{8c_s(1 + \frac{N^2 - 4}{16}\gamma^2)^2}{(N^2 - 4)\gamma} \quad , \quad    \tilde{\psi}(\delta) = -\cfrac{{V_*{(1+\delta^2}})^{2}}{2 \delta} \label{app:pdeltaOgden}          
\end{align}
Substituting the above expressions in \eqref{app:dtmdalphatwo} we obtain
\begin{align}
\tilde{G}(\delta) &= 1 + 2\lambda_M - \frac{\lambda_M(1+\delta^2)}{2\delta^2}  -\cfrac{{3{(1+\delta^2}})^{2}}{2\delta^2(3+\delta^2)}\frac{(n-1)\lambda_M}{n}\\
&= 1 + \frac{3\lambda_M}{2} - \frac{\lambda_M}{2\delta^2}  -\cfrac{{3{(1+\delta^2}})^{2}}{2\delta^2(3+\delta^2)}\frac{(n-1)\lambda_M}{n}
\end{align}

\noindent Using \eqref{app:c_ogden} and \eqref{app:pdeltaOgden} in \eqref{specialisedXmGent} we have 
\begin{align}
X_{M} &= \frac{\lambda_M t_0}{n V_0^\frac{1}{n}} \eval{\frac{\tilde{\psi}\tilde{c}}{(-\tilde{Q}(\delta))^{\frac{n-1}{n}}}}_{\delta_{\rm{min}}^M} = \frac{c_s t_0 V_*  \lambda_M}{n V_0^{\frac{1}{n}}} \eval{ \cfrac{-(1+\delta^2)^3}{2\delta (-Q(\delta))^{\frac{n-1}{n}}} }_{\delta_{\rm{min}}^M}  \\
&= \frac{c_s t_0 V_*^{\frac{1}{n}}  \lambda_M}{n V_0^{\frac{1}{n}}} \eval{ \cfrac{K_\textrm{og}(\delta)}{(-q(\delta))^{\frac{n-1}{n}}} }_{\delta_{\rm{min}}^M}  = \frac{L_*\lambda_M}{n m^{\frac{1}{n}}}\eval{ \cfrac{K_\textrm{og}(\delta)}{(-q(\delta))^{\frac{n-1}{n}}} }_{\delta_{\rm{min}}^M} \label{app:ogdenresult1}
\end{align}    
Plugging \eqref{deltamingent2} in \eqref{app:ogdenresult1} we obtain
  \begin{equation}
\frac{X^M}{L^*} = \cfrac{\lambda_M}{nm^{\frac{1}{n}}}\begin{cases}
\cfrac{K_{\rm{og}}(-\delta_0(\lambda_M,n))}{ m_{\rm{th}}^\frac{n-1}{n}}  & m \ge m_{\rm{th}}, n>0.5 \\[15pt] 
\eval{\cfrac{K_{\rm{og}}}{(-\tilde{q})^\frac{n-1}{n}}}_{-\tilde{q}^{-1}(m)}  & m < m_{\rm{th}}, n > 0.5 \\
 \end{cases} 
\end{equation} 

\section{Post first shock formation}
\label{app:postshock}
\noindent Consider the loading of an initially undeformed strain stiffening solid by the load \eqref{loadprofile} so that we have shocks. Let us assume that after a sufficiently long time from the first shock formation, a steady state pure shock propagates into the material, i.e there is a propagating velocity and strain discontinuity, with a velocity jump from $0$ to $V_0$ and a strain jump from $0$ to $\gamma^-$ across the shock. The momentum balance equation \eqref{momentum_conserv_integ_Lagrangian} and the integral form of \eqref{compatibility}, when applied to the case of a propagating pure shock, give us the jump conditions
\begin{equation}
\dot s \jump{\gamma} + \jump{v} = 0 \quad , \quad \jump{\tau} + \rho_0 \dot s \jump{v}= 0 \label{dummy_postshock}
\end{equation}
where, $\dot s$ is the shockspeed and $\jump{f} = f^+ - f^-$ denotes the jump in a field $f$ across a shock, from its value $f^+$ ahead of the shock to $f^-$ behind it. Eliminating $\dot{s}$ in \eqref{dummy_postshock} we obtain
\begin{equation}
\jump{\tau}\jump{\gamma} = \rho_0 \jump{v}^2 \label{jump3}
\end{equation}
Using $\gamma^+ = \tau(\gamma^+) = v^+ = 0$, and $v^- = V_0$ in \eqref{jump3}, and defining the auxiliary  function $\phi(\gamma) = \tau(\gamma)\gamma$ we have
\begin{equation}
\phi(\gamma^-) = \rho_0 V_0^2 \label{app_dummy3_postshock}
\end{equation}
Consider cases in which $t_\textrm{shock}>t_0$, that is the first shock does not form before the ramping is completed. Then the largest strain (in magnitude)  realized before first shock formation, $\gamma_1$, would be the strain state at the trailing edge of the waveform where the velocity is equal to $V_0$, and is obtained from (using $\eqref{Qsol}$, $\eqref{Q_def}^1$, and definition of $c$ in $\eqref{waveeqn}$)
\begin{equation}
Q(\gamma_1) = \int\displaylimits_0^{\gamma_1} \sqrt{\frac{\tau'(\gamma)}{\rho_0}} \dd{\gamma} = -V_0  \label{app_dummy_postshock}
\end{equation}
Squaring \eqref{app_dummy_postshock} and defining the non-dimensional auxiliary function $\chi(\gamma) = (\int\displaylimits_0^\gamma \sqrt{\tau'(\gamma)} \dd{\gamma})^2$ we obtain
\begin{equation}
\chi(\gamma_1) = \rho_0 V_0^2  \label{app_dummy2_postshock}
\end{equation}
For a general strain stiffening solid the functions $\chi(\gamma)$ and $\phi(\gamma)$ are different, and hence, from \eqref{app_dummy3_postshock} and \eqref{app_dummy2_postshock} the strain state at the trailing edge of the evolving waveform, $\gamma_1$, is different from the strain state behind a steady state shock. As mentioned in Sec. \ref{sec:exponential}, when applying the method of characteristics solution \eqref{sol} behind the shock, $f(\beta)$ is not constant and the $\beta$ characteristics come into play, meaning that there is also information traveling back to the loading surface from the growing shock. This might facilitate the change in strain state from $\gamma_1$ to $\gamma^-$ for possible steady state shock propagation but this is subject for future research. See Ch. 4 in \cite{hamilton1998nonlinear} and the references therein for a discussion on reflection of waves by shocks in gas dynamics. There, it is demonstrated that for weak shock propagation the ``reflected wave'' is extremely small and can be neglected.  
%
%
%


\end{document}